\def \newpage {%
  \if@noskipsec
    \ifx \@nodocument\relax
      \leavevmode
      \global \@noskipsecfalse
    \fi
  \fi
  \if@inlabel
    \leavevmode
    \global \@inlabelfalse
  \fi
  \if@nobreak \@nobreakfalse \everypar{}\fi
  \par
  \vfil
  \penalty -\@M}
\documentclass[amsmath,amsfonts,amssymb,twocolumn,superscriptaddress,showpacs]{revtex4-1}

\usepackage{graphicx,psfrag,color}
\usepackage{dcolumn}
\usepackage{bm}
\usepackage{mathbbol, appendix}
\usepackage{multirow}
\usepackage{epsf}
\usepackage{epsfig}

\begin{document}

\draft

\title{Topological Charge Order and Binding in a Frustrated XY Model
and Related systems}
\author{Zohar Nussinov}
\affiliation{Department of Physics, Washington University, St.
Louis, MO 63160, USA}

\begin{abstract}
We prove the existence of a finite temperature $Z_{2}$ 
phase transition for the topological charge 
ordering within the Fully Frustrated XY Model.
Our method enables a proof of the topological 
charge confinement within the conventional 
XY models from a rather 
general vista. One of the 
complications that we
face is the non-exact 
equivalence of the continuous (angular) 
XY model and its discrete topological charge dual.
In reality, the energy spectra 
of the various topological 
sectors are highly nested
much unlike that suggested
by the discrete dual 
models. We surmount these difficulties
by exploiting the 
Reflection Positivity
symmetry that this periodic 
flux phase model possesses.
The techniques introduced
here may prove binding 
of topological charges in numerous
models and might be applied
to examine transitions
associated with various
topological defects,
e.g., the confinement
of disclinations
in the isotropic 
to nematic transition. 
\end{abstract}

\maketitle

\section{Introduction}

In this work, we examine the topological
charge ordering of XY models in the 
presence of various external  
magnetic fields and prove, by Reflection
Positivity, that these charges 
order at finite temperature 
in the ``Fully Frustrated XY Model'' 
(wherein, as we will explain, half a fluxon threads
each elementary plaquette in an XY model). Beyond the particular quite specific
result related to transitions in the fullly frustrated XY (and other spin) models \cite{teitel-review,choi,domany,knops,early1,early2,early3,early4,early5,ettore,ettore'}, the principal
ideas that we advance and the reflection positivity techniques that
we employ \cite{oldme} might be used to prove topological charge order
in a host of systems including those with continuous symmetries for
which discrete Coulomb type and other representations of these charges 
are highly illuminating yet approximate
and may lead to incorrect results. Offshoots of the methods that we invoke demonstrate
a relatively universal attraction between 
opposite charges in these (and many other symmetric (or, 
more precisely, ``Reflection Positive'' \cite{RP})  
systems. Extensions of the techniques employed here have
ramifications in other fields such as
the sign of Casimir effects, e.g., \cite{klich}. 
Questions concerning the existence of
topological charge order in frustrated systems
have, in recent years, gained impetus from
the study of topological quantum orders \cite{TQO1,TQO2,TQO3,jaffe}.
The study of uniformly frustrated systems have also been of interest in 
study of glassy dynamics and aging \cite{Halsey,me,tarjus,walter}. 

The outline of this work
as follows. In section (\ref{Defn})
we define the two dimensional frustrated XY model.
In section (\ref{approximate}), a simple
set of general trigonometric difference equations 
and approximate solutions for ground 
state configurations will be given.
Much of this section is not new
but is merely our perspective
on the matter. We
illustrate how in the 
{\em low-energy sector} the 
continuous XY model and 
its discrete topological charge dual 
are almost one and the 
same.  In section(\ref{exact}),
several simple exact 
ground state solutions
are given. We then compute energy gaps 
between different topological 
charge configurations which will
serve as
a springboard in our proof
of finite temperature phase 
transitions.

In the sections thereafter, we largely focus on new non-trivial results.
In section(\ref{FFXY}), we investigate the Fully
Frustrated XY model. Employing the results of section (\ref{exact}),
in subsection(\ref{gaps}), we establish the existence of energy 
gaps between the different topological 
charge sectors. We then show, in subsection (\ref{RP}), 
that the system is Reflection Positive and
consequently prove a finite temperature
topological charge ordering transition.
In section(\ref{altern}) we
provide an alternative
proof to the ordering 
transition.
We demonstrate, in section(\ref{stand}), that by a 
minor variation, our proof immediately 
relates to the usual finite temperature 
annihilation of vortex-antivortex pairs 
occurring in the Kosterlitz-Thouless
transition. 
In section(\ref{kaleid}) 
we discuss systems with kaleidoscope 
pattern frustrations. 
In section (\ref{Z_k}), we discuss the extension
of the chiral $Z_{2}$ symmetry of the Fully 
Frustrated XY Model to a $Z_{k}$ symmetry
present for other rational  frustration $f$
which tends to 
an $O(2)$ symmetry for irrational $f$. Numerically, with the Fully Frustrated XY model,
both symmetries 
(i.e., (i) the $Z_{2}$ symmetry which as we prove in this work is broken at low temperatures and (ii) the continuous $O(2)$ symmetry)
are lifted at very close temperatures
 ($T \simeq 1.286 J$ (with $J$ the 
exchange constant)) \cite{teitel-review,teitel-reply}.
We suggest an analogy to the 
magnetic groups appearing 
in the Quantum Hall 
problem.
We end, in section(\ref{nematic}),
in an application
of these ideas to other
topological defects-
e.g. disclinations
in liquid crystals.


\section{Definition of 
the Model}
\label{Defn}

An excellent review of the Fully Frustrated XY Model
is provided in \cite{teitel-review}. We proceed with a particular common rendition of this model , e.g., \cite{tj,tj'} which we will henceforth analyze. 
We consider a classical XY model on
an $L \times L$ square lattice with periodic 
or open boundary conditions (with $2L^{2}$ 
or $2 L^{2}+2L$ bonds 
respectively). The two component
spins $\{ \vec{S}_{i} \}$ at each
lattice site $i$ are labeled
by their orientation $\{\theta_{i}\}$ relative
to a chosen axis.
The Hamiltonian
\begin{eqnarray}
H = - J \sum_{\langle i j \rangle} \cos(\theta_{i}-\theta_{j}-A_{ij})
\label{definition}
\end{eqnarray}
where $i,j$ are nearest neighbor sites on the 
lattice. Throughout this work, unless stated otherwise,
we set both the exchange constant $J$,
and (as seen above) the lattice constant 
to one. Two symmetries of
this Hamiltonian
are evident:

$\bullet$ A global $O(2)$ symmetry. 
The transformation $\theta_{i} \to \theta_{i} + \phi$
~ for all sites $i$ with a uniform shift $\phi$ leaves the Hamiltonian $H$
trivially invariant. By the Mermin-Wagner-Coleman \cite{Mermin-Wagner,Coleman}
theorem, continuous 
symmetries are never broken in 
two dimensions:  at any finite temperature the local
order parameter $\langle \theta_{i} \rangle=0$
for all twice differentiable,
rotationally symmetric, momentum space 
interaction kernels \cite{me}.
In this two dimensional system, the superfluid
phase stiffness exhibits a first order jump 
at a Kosterlitz-Thouless 
temperature $T_{KT}$ below which rotational
symmetry is  (marginally) ``broken'' to produce a state
with algebraic long range order
\cite{KT,Ber,minnhagen,minnhagen'}.

$\bullet$ For the special cases in
which the flux $\Phi_{R} \equiv \sum_{\Box R} A_{ij}$
is the same for all plaquettes $R$, 
translational invariance is also
a symmetry of the Hamiltonian.
We shall primarily focus (for the 
Fully Frustrated XY Model) on the breaking of 
translational invariance under a shift
by one lattice constant, 
$\theta_{i} \rightarrow \theta_{i+\hat{e}_{1}}$
for all $i$, as seen by the arrangement of vortices.
A critical temperature $T_{vortex}$ may be associated with 
the breaking of this symmetry.  
Numerically, as the temperature is lowered within the FFXY, 
it is found that $|T_{vortex}-T_{KT}| \ll T_{KT}$. 
Whether the two temperatures might be
exactly one and the same has been a puzzle
surrounded by some controversy,
albeit very promising recent work
\cite{kink-new}.

Eq. (\ref{definition}) describes an array of 
Josephson junctions immersed in an external magnetic
field, e.g., \cite{teitel-review}. It may model an extreme type II 
superconductor built of discrete 
superconducting elements (``grains'') 
at the various lattice sites where it is 
assumed that the modulus of the superconducting 
wave-function on each grain is pinned and that  
only its phase which may vary  \cite{continuum}.
The evolution of the phases is
easily gleaned within a path integral 
formulation: an external 
electromagnetic vector 
potential $A_{\mu}(\vec{x})$
gives rise to the differential
action $dS_{em} = - \frac{e^{*}}{c} A_{\mu} dx^{\mu}$
for a displacement $dx^{\mu}$, with $e^{*}$
the effective charge and $c$ the speed
of light. This
leads to a phase shift
$\exp[i S_{em}/\hbar]  \rightarrow \exp[i A_{ij}]$
along the link $\langle ij \rangle$
where
\begin{eqnarray}
A_{ij} =  - \frac{e^{*}}{\hbar c} \int_{\vec{i}}^{\vec{j}} \vec{A} \cdot \vec{dx}.
\end{eqnarray}

It follows that in going around a plaquette 
the electromagnetic field incurs an 
additional phase 
\begin{equation}
  - \frac{e^{*}}{\hbar c} \oint_{R} \vec{A}(\vec{x}) \cdot \vec{dx}
= \sum_{\Box R} A_{ij},
\end{equation}
the elegant Aharonov-Bohm effect \cite{AB}.
The above directed loop sum may be replaced
(by a simple application of Stokes theorem)
by the magnetic flux threading the
plaquette to conform with the 
conventional experimental prognosis
of the effect. The elementary plaquette $R$ is said to be frustrated 
if the Aharonov-Bohm 
phase associated with it is not congruent 
to zero. In such an instance, the net magnetic flux piercing
$R$ is not an integer multiple of the elementary fluxon $\Phi_{0}$.
If the directed sum is defined as 
\begin{eqnarray}
\sum_{\Box R} A_{ij} \equiv 2 \pi f_{R},
\label{f}
\end{eqnarray}
then the ``frustration'' $f_{R}$
is not an integer in such a
case.
 
For the type II superconductor, 
$f = \frac{a^{2} B}{\Phi_{0}}$
with $a$ the lattice constant, $B$ the 
magnetic field strength, and 
$\Phi_{0} = \frac{h c}{2 e}$
the elementary magnetic fluxon
($e^{*}= - 2e$ for the Cooper pairs).  
All of the above may be formulated in a 
gauge independent manner. We define
$\phi_{ij} \equiv \theta_{i}-\theta_{j} - A_{ij} ~~(mod~2 \pi)$ 
such that $-\pi < \phi_{ij} \le \pi$.
The Hamiltonian is now a function 
of $\{ \phi_{ij} \}$.
Unimportant gauge degrees freedom
have been absorbed. In a 
system with periodic boundary conditions,
we are left with $N=L^{2}$ constraints 
of net circulation, one
for each plaquette $R$: 
\begin{eqnarray}
\{~ \sum_{\Box R} \phi_{ij} = 2 \pi (m_{R}-f_{R})~ \}
\label{ilutz}
\end{eqnarray}
with integer $\{ m_{R} \}$ \cite{degs_of_freedom}.
The content of the 
``vorticity'' $m_{R}$ is clear.
Let us slowly trace its 
origin and justify 
the latter name.  
The vorticity of a 
plaquette $R$:
$m_{R} = \sum_{ \langle ij \rangle \in R} m_{ij}$
where $\phi_{ij} = \theta_{i}-\theta_{j}-A_{ij}+ 2 \pi m_{ij}$.
The vorticity $m_{R}$ represents the multiple
of $(2 \pi$ ) that needs to be added to the 
bare gauge invariant angular differences 
$(\theta_{i}-\theta_{j}-A_{ij})$
such that $-\pi <\phi_{ij} \le \pi$. As such, 
it counts the number of times that
the XY rotor circulates as we go 
counterclockwise round the plaquette $R$.
We can immediately make
two statements 
regarding
the effect of a 
spatially uniform frustration $f$: 

(i) The 
ground state energies are trivially
continuous in $f$. The proof is 
very simple: By Eq.(\ref{definition}), we note
that for each 
given angular configuration $\{ \theta_{i} \}$,
the energy $E(\{ A_{ij} \}; \{ \theta_{i} \})$ 
is a continuous function of $\{A_{ij}\}$; 
each individual gauge link variable $A_{ij}$
appears as an argument of single continuous
cosine function. Thus the ground state energy (i.e. 
the minimum amongst all of these continuous 
energy curves, each corresponding to a different angular 
configuration $\{\theta_{i}\}$), $E_{ground} = \inf_{\theta} 
\{E(\{ A_{ij} \}; \{ \theta_{i} \})\}$
is also a continuous function of  $\{A_{ij}\}$ 
or, in a gauge invariant formulation, the ground state
energy per link is a continuous function of the 
frustration $f$ \cite{scaling}. This 
continuity argument is shorter than that spanned by 
an entire earlier Letter \cite{E_f}. 

(ii) The system is invariant
under the transformations
\begin{eqnarray}
f \rightarrow n \mp f
\label{lanl}
\end{eqnarray} 
for all integers $n$.
Once again, the proof is trivial:
the inversion of the external 
magnetic field
$f \rightarrow -f$ 
along with $\theta_{i} \to - \theta_{i}$
leaves the partition function
$Z = \prod_{i} \int d\theta_{i} ~\exp[-\beta H]$
identically the same 
as the cosine interaction
term is an even function.
The periodicity of the cosine 
secures an additional
invariance: 
$f \rightarrow n+f$ \cite{modular}.
Compounding the last
two operations
together proves the symmetries of Eq.(\ref{lanl}).
These imply that, 
as a function of external magnetic
flux,  the ground state
energy $E_{ground}(f)$ has local extrema
at the symmetry points 
$f=0, \pm 1/2, \pm 1, ...~$.
The points $f \equiv 0$ (mod 1) are 
global minima of  $E_{ground}(f)$-
all bonds are saturated at their minimal
value $(-1)$ within the ground state.

The symmetries of Eq.(\ref{lanl}) have led 
the $f=1/2$ variant to be regarded as
the ``Fully Frustrated" one. 
After all, any larger value of 
$f$ may be folded back to another
value of $f$ lying in the interval $[0,1/2]$;
the model with $f=1/2$ is as frustrated 
as it can possibly get. 
However, in some sense, the ``Fully Frustrated XY model"
is the least frustrated of
all. The time reversal symmetry
(or, more formally, a Reflection Positivity 
invariance) that this model possesses
will allow us to prove
that it displays a 
$Z_{2}$ phase transition.

\section{Approximate Ground States and The Low Energy 
Dual (topological charge) Model}
\label{approximate}

\begin{figure}[htbp]
 \begin{center}
     \epsfxsize0.52\columnwidth
\epsfbox{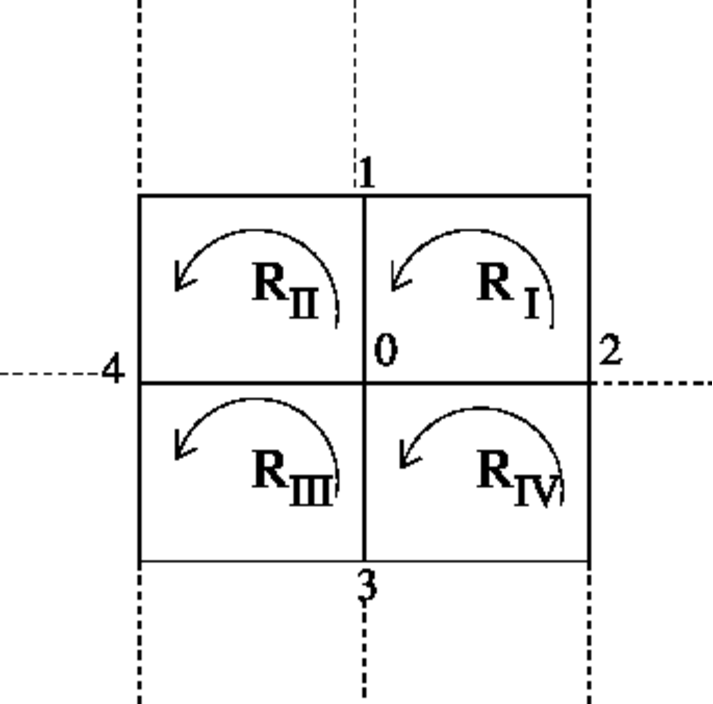}
\caption{A lattice fragment
centered about the origin.}
 \label{fig:Laplacian}
\end{center}
\end{figure}

The current section, unlike
others to follow, is not 
rigorous. Much of its
contents are not new
(albeit often derived
by new routes). Its intent 
is to provide the 
reader with better
intuitive grip
on the physics of
the frustrated XY model.
Experts may choose to directly study
Sections \ref{exact} or \ref{FFXY}. 
Enforcing the constraints of given 
vorticities, Eqs.(\ref{ilutz}), the 
Hamiltonian to be 
minimized reads
\begin{eqnarray}
{\cal{H}}= - \sum_{ \langle ij \rangle} \cos \phi_{ij} 
+ \sum_{\alpha} \lambda_{\alpha} \sum_{\Box R_{\alpha}} \phi_{ij},
\label{general}
\end{eqnarray}
where $\{\lambda_{\alpha}\}$ 
are Lagrange multipliers.
Unless some of the 
angles lie on the boundaries
($\phi= \pm \pi$), at the extrema 
\begin{eqnarray}
\sin \phi_{ij} + \lambda_{ij \in \alpha} - \lambda_{ij \in
\alpha^{\prime}}=0
\label{minimum}
\end{eqnarray}
where the two adjacent plaquettes 
$R_{\alpha},R_{\alpha^{\prime}}$ 
share the bond $\langle ij \rangle$. 
The Lagrange multipliers$\lambda_{ij \in \alpha}$ and  
$\lambda_{ij \in \alpha}^{\prime}$
refer to the plaquettes $\alpha$ and $\alpha^{\prime}$
which share the same bond $\langle ij \rangle$ (one having 
it appear in a counterclockwise circulation
of $\alpha$ and the other having it 
in a clockwise reading of 
$\alpha^{\prime}$).  
Examining Fig.(\ref{fig:Laplacian}) we see that 
\begin{eqnarray}
\sin \phi_{01} = \lambda_{I}-\lambda_{II},~ \sin \phi_{04} = \lambda_{II} - \lambda_{III} \nonumber
\\ \sin \phi_{03} = \lambda_{III} - \lambda_{IV}, ~ \sin \phi_{02} = \lambda_{IV} - \lambda_{I}. 
\end{eqnarray}

Summing it all up,
\begin{eqnarray}
\sum_{\langle 0 j \rangle} \sin \phi_{i=0,j} =0,
\label{node}
\end{eqnarray}
the sum of all Josephson 
currents entering 
any node 
must vanish.
We also arrive at Eq.(\ref{node})
when minimizing 
Eq.(\ref{definition})
within a given 
gauge to find the 
global ground state. 
Eqs.(\ref{general},\ref{node})
show that current conservation 
holds at extrema 
of {\it each of the
individual topological
sectors} $(m_{1}, m_{2}..., m_{N})$
(unless the minima
lie on the boundaries
of the topological sectors:
i.e. $\phi_{ij} = \pm \pi$). 

If $f=0$ and $m_{R}=0$ in the region
outside a given set of vortices 
then we may work in a gauge in which 
$\phi_{ij} = \theta_{i} -\theta_{j} \equiv \theta_{ij}$.
In this gauge, Eq.(\ref{node})
reads 
\begin{equation}
\nabla_{\mu} (\sin [\nabla^{\mu} \theta])  = 0,
\label{ZN}
\end{equation}
with $\nabla^{\mu}$ denoting 
the discrete lattice 
difference along the 
$\mu$ direction.
This is equivalent to the difference equations
\begin{equation}
\nabla^{\mu} \theta  = \sin^{-1}
\Big([\vec{\nabla} \times
 \vec{{\cal{A}}}(\vec{x})]^{\mu} \Big).
\end{equation}
  The function $\vec{{\cal{A}}}(\vec{x})$
must satisfy
\begin{eqnarray}
\sum_{bonds \in \Gamma} \theta_{ij} =  \sum_{\langle ij \rangle \in
\Gamma} 
\sin^{-1}
\Big([\vec{\nabla} \times
 \vec{{\cal{A}}}(\vec{x})]^{\mu_{ij}} \Big)
\end{eqnarray}
for all contours $\Gamma$.
Here $\mu_{ij}$ are chosen to lie
along each of the individual 
bond directions ${\langle i j \rangle}$ 
in the contour 
product.
In the continuum limit
Eq.(\ref{ZN}) becomes
\begin{eqnarray}
\cos^{2} \frac{\theta}{2} \partial_{\mu}  \partial^{\mu} \theta(\vec{x})
\equiv D^{2} \theta(\vec{x})=0. 
\label{Lap}
\end{eqnarray}
Effectively, the phase $\theta(\vec{x})$ isotropically 
dilates the metric.
Apart from the special case $\theta = \pm \pi$, 
we obtain the Laplace equation. Thus, in the vortex free ($m=0$),
frustration free ($f=0$), region $\theta(\vec{x})$ is 
a harmonic function. 
Let us now return to the discrete lattice equation Eq.(\ref{ZN})
and examine its behavior in the linear regime far away from 
the vortices when the angular differences are small 
we can replace Eq.(\ref{ZN})
by $\nabla_{\mu}\nabla^{\mu} \theta=0$.
Viewing the Josephson current as a fictitious magnetic field
$\vec{{\cal{B}}} = \vec{\nabla} \theta$
(with $\vec{\nabla}$ the discrete lattice gradient),
the condition for a minimum becomes
$\vec{\nabla} \cdot \vec{{\cal{B}}} =0$ 
at all lattice sites.
On the other hand, the boundary 
conditions (recall that we 
focused on a region in which 
$m_{R}=f_{R}=0$ such that a decomposition
$\phi_{ij} = \theta_{i}-\theta_{j}$is possible)
imply that
\begin{eqnarray}
\sum_{C} \vec{{\cal{B}}} \cdot \delta \vec{l} = 2 \pi [m_{net}-f_{net}]
\end{eqnarray}
where $m_{net}$ and $f_{net}$ are respectively the net topological
charge and external magnetic flux/$(2 \pi)$ in the 
region bounded by the contour $C$.
Choosing a gauge for the pseudo magnetic 
field $\vec{{\cal{B}}} = \vec{\nabla} \times \vec{{\cal{A}}}$
such that $\vec{\nabla} \cdot \vec{{\cal{A}}}=0$
we obtain an equation similar, 
in the thermodynamic limit,
to the lattice version of the 
standard Biot-Savart law
\begin{eqnarray}
\vec{{\cal{A}}}(\vec{x}) = 2 \pi \sum_{x^{\prime}}
[m_{\vec{x}^{\prime}}-
f_{\vec{x}^{\prime}}] G(\vec{x},\vec{x}^{\prime}) \nonumber
\\ G(\vec{x},\vec{x}^{\prime}) = \int_{B.Z.} \frac{d^{2}k}{(2 \pi)^{2}} 
~~\frac{\exp[i \vec{k}
\cdot (\vec{x}-\vec{x}^{\prime})]}{4 -2 \cos k_{1}- 2 \cos k_{2}},
\end{eqnarray}
where $m_{\vec{x}}$ and $f_{\vec{x}}$ are the local
vorticity and frustration in the plaquette $\vec{x}$.
The sum over $\vec{x}^{\prime}$ spans the entire lattice.
(In higher dimensions the denominator readily 
generalizes to $2\sum_{l=1}^{d} (1- \cos k_{l})$.) 
Employing the ``Biot-Savart'' law 
and the magnetic analogy,
boundary effects may be included by 
extending the integral to include 
images of the singularities.
The ground state energy given a certain 
distribution of $\{m_{\vec{x}}\}$ and $\{f_{\vec{x}}\}$ outside
our domain is 
\begin{eqnarray}
E_{ground} = \frac{1}{2} \sum_{\vec{x},\vec{x}^{\prime}} 
[m_{\vec{x}}-f_{\vec{x}}] G(\vec{x},\vec{x}^{\prime}) 
[m_{\vec{x}^{\prime}}-f_{\vec{x}^{\prime}}].
\label{topo1}
\end{eqnarray}
The ``Coulomb gas'' type character of the XY model suggested by Eq. (\ref{topo1}) 
is well-known and has been derived by various methods both approximate 
\cite{teitel-review,villain1977,coulomb1,coulomb2,coulomb3}
and explicitly related to more exact duaities \cite{bond,clock}.
The Coulomb gas representation has played a commanding role in the understanding
of the frustrated (and non-frustrated) XY model.
A principal complication that we will face in our endeavor of establishing low temperature
transitions is that the typical dual Coulomb gas representation
is {\it not exact} and thus may give rise to incorrect results. The techniques that 
we use in the current work allow us surmount these difficulties and will enable us prove the existence of low temperature topological charge order associated with the 
vorticities $m_{\vec{x}}$.

In the small $k$ (continuum) limit , the kernel $G(\vec{x}-\vec{x}^{\prime})$ above becomes
logarithmic as befits the two-dimensional Coulomb charges. The resulting expression
becomes identical, for all practical purposes, 
to the Villain model \cite{villain1977}. 
In this limit, the energy 
of interaction between two vortices is
\begin{eqnarray}
E = 2 \pi q_{1} q_{2} \ln (a/r_{12}).
\label{topo2}
\end{eqnarray}
where $q_{i} \equiv (m_{i}-f_{i})$. 
For the time being,
we have inserted a length
``$a$'' in order to make
the argument of the 
logarithm dimensionless.
When supercurrent 
screening is introduced, ``$a$'' 
naturally becomes the 
screening length in 
the short distance limit.
Thus, in this unscreened case,
``$a$'' may be regarded
as the system size:
$a = O(L)$ \cite{screening}.

In the continuum limit lines connecting 
opposite sign vortices become cuts in the 
complex plane with branch points at the positions
of the vertices.  A solution to the Laplace
equation (Eq.(\ref{Lap})) jumping by $2 \pi q_{i}$ across the 
branch cut while traversing each pole
(vortex) $z_{i}= x_{i~1}+ i x_{i~2}$
(where $(x_{i~1},x_{i~2})$ are
the vortex coordinates within the
plane)  is the well known \cite{Polyakov}
\begin{eqnarray}
\theta(z) = \sum_{i} q_{i} Im \{ \ln(z-z_{i}) \}.
\end{eqnarray} 
In the previous (magnetic)
analogy the singular branch cut
takes on the role
of a Dirac like string.
If we have a pair
of equal and opposite vortices $q_{1} = -q_{2}$, 
then $\theta/q_{1}$ simply becomes the
``angle of site'' of the segment
connecting $z_{1}$ and $z_{2}$
as viewed from the point $z$.
$\theta$ may be written
as the phase of wave-function (which may be superficially
viewed as a Laughlin-like wavefunction \cite{laughlin} as a function of 
only one of the ``electronic'' coordinates) with the odd integer
power replaced by $q$ \cite{Laugh}: 
\begin{eqnarray}
\psi(z) = \prod_{i} (z-z_{i})^{q_{i}}.
\label{Laugh}
\end{eqnarray}
The phase of this wavefunction
$\theta =   Im \{ \ln \psi(z) \}$.

The continuum limit unfrustrated XY Hamiltonian,
\begin{eqnarray}
H = \frac{1}{2}  \int d^{2}x (\nabla \theta)^{2} 
\nonumber
\\ = \frac{1}{2}  \int d^{2}x
\vec{\nabla} Im \{
\ln \psi(z) \}
\cdot
\vec{\nabla} Im \{ \ln \psi(z) \} \nonumber
\\
= \frac{1}{2}  \sum_{i,j} q_{i}q_{j} \int   d^{2}x
\vec{\nabla} Re \{
\ln(z-z_{i}) \} 
\cdot
\vec{\nabla} Re \{ \ln (z-z_{j}) \},
\nonumber
\end{eqnarray}
where we employ the Cauchy-Riemann relations 
\begin{eqnarray}
\partial_{\alpha} u = \epsilon_{\alpha \beta} \partial_{\beta} v
\Longrightarrow ~ 
(\vec{\nabla} u)^{2} = (\vec{\nabla} v)^{2}
\end{eqnarray}
(with $\epsilon_{12} = -\epsilon_{21}=1$)
for the complex function $ F(z) = \ln \psi(z) \equiv u + i v$
(where the squared gradients are
$1/r^{2}$). Integrating by parts,
\begin{eqnarray}
H = -\frac{1}{2} \sum_{i,j} q_{i} q_{j} \int d^{2}x \ln|z-z_{i}|
\nabla^{2} 
\ln|z-z_{j}|.
\end{eqnarray}
By the $2 \pi$ discontinuity around 
a logarithmic pole,
$\nabla^{2} \ln|z-z_{j}| = 2 \pi \delta(z-z_{j})$.
Thus,
\begin{eqnarray}
H = -\pi  \sum_{i,j} q_{i} q_{j} \ln|z_{i}-z_{j}| \nonumber
\\ =  \pi  \sum_{i,j} q_{i} q_{j} \ln (a/r_{ij}) + const
\label{topo3}
\end{eqnarray}
with an arbitrary $a$ and inter-vortex separation $r_{ij} = |z_{i} - z_{j}|$.
This swift independent derivation 
is merely a slight twist on 
the treatment of Itzykson and Drouffe
\cite{ID}. Note that the final result is 
identical to Eq.(\ref{topo2})
derived by a seemingly alternate 
route. Such a plasma like 
interaction is also present 
in the Boltzmann weights 
of Quantum Hall Laughlin states. 
Along this route, the standard Villain model duality \cite{villain1977} 
acquires a new meaning-
an interchange between the 
real and imaginary parts
of a complex function $F(z)$.
Here the nature of the duality
is manifest: the imaginary part
of $F(z)$ leads to a description
in terms of the angular configuration
of the XY rotors. The real part of $F(z)$
leads to a description in terms of the 
topological charges of the defects
(when the frustration $f=0$). Formally, 
we linked the two dual pictures by 
employing the Cauchy-Riemann 
relations. We note that for
$f=1/2$, the $Z_{2}$ transition
temperature $T_{vortex}$
corresponds to the 
ordering of charges 
in the dual model,
whereas $T_{KT}$
corresponds to
the critical temperature
in the original
angular variables. 
Number (charge) and phase variables 
appear in many instances (e.g. 
in the  
triangular XXZ models displaying 
both $Z_{2}$ and XY
transitions
with nearly equivalent
temperatures \cite{triangle-XXZ}).

We now provide a new geometrical interpretation of 
the frustrations.
Forgetting about boundary conditions, let us arrange
the lattice sites on a sphere of radius $R$ (or many fractured
portions of such spheres pasted together). 
The sum of the angles of a spherical 
quadrilateral $= 2 \pi$ + Area$/R^{2}$. It
follows that in this naive mapping, we can 
identify a uniform frustration $f=$ (Area of the 
plaquette) $/R^{2}$.  
The frustrated XY Hamiltonian is now 
\begin{eqnarray}
H = - \sum_{ \langle i j \rangle} \vec{S}_{i} \circ \vec{S}_{j}
\end{eqnarray}
where $\circ$ denotes the scalar product between 
the vector resulting from parallel
transporting the spin $\vec{S}_{i}$ to site $j$
and the spin $\vec{S}_{j}$ 
already situated at site $j$. 
This Hamiltonian is translationally invariant
when $f =const$ and 
may be diagonalized in Fourier space
\cite{me,Non-Abelian}. By this crude analysis, we would
expect the ground state modulation 
length (for a constant value
of $f$) to roughly scale as the radius $R \sim f^{-1/2}$
for small fractions $f$
such that one fluxon 
pierces the surface of the sphere
(giving rise to a trivial Aharonov 
Bohm phase). This conforms with  
numerical observations for $f=1/q$ with an integer 
$q \gg 1$: each fluxon (or vortex) occupies an 
area of size $1/f$
\cite{Straley}.

More conventionally, requiring overall charge neutrality
in the Villain model (or for our
pseudo-currents $\{q_{a}\}$ that interact
Coulombically) yields the same result. 
By net charge neutrality, the background charge $ [-\sum_{a} f_{a}]$
should cancel against $[\sum_{a} m_{a}]$
originating from the vorticities. 
Once again, from this ``angle'' as 
well, the density  of vortices should
scale as $f^{1/2}$ for 
a small uniform frustration 
$f$. In any system having periodic boundary
conditions, the net topological charge 
must vanish. It follows that any system 
having ferromagnetic boundary 
conditions cannot enclose a net nonzero 
vorticity. All vortices must appear in 
vortex-antivortex pairs. On a less 
precise level, within the
approximate dual model, this 
follows from a divergent Coulomb 
penalty for any lone unpaired 
vortex. Let us define a radius $r$
vortex in the unfrustrated problem 
as one in which 
\begin{eqnarray}
\sum_{\langle ij \rangle \in \Gamma} \phi_{ij} = 2 \pi m_{R},
\label{ar-vortex} 
\end{eqnarray}
with $m_{R} = 1$ is satisfied for all contours $\Gamma$ which 
encircle the plaquette $R$ up to a maximal 
distance $r$ away from it. Employing the 
Euler-Maclaurin summation formula,
we find that for large $r$, the energy 
of a radius $r$ vortex is bounded from 
below by that of a vortex in the 
continuum limit where an integration
may be performed. In the large $r$
limit the energy may bounded from below 
by $K_{1} \ln (r/r_{1})$ where, in this
unfrustrated case, the constants 
$0< K_{1} < \pi$ and $r_{1} > 1$.
``Phase space'' volume calculations
may also be enacted for these systems \cite{pv}.

\section{Exact Flux Patterns}
\label{exact}

All that follows in this section
and hereafter is new and exact and is not
based on any of the approximations 
of section (\ref{approximate}). 
The periodicity (if any) of the 
ground state phase pattern $\{ \theta_{i} \}$ 
depends upon the gauge.
Even in the unfrustrated ($f=0$) case
a gauge with irrational $\{A_{ij}\}$ 
may be chosen such that
the phases are never periodic.
As throughout most of this paper
we will attempt to focus on the 
gauge invariant bond angles 
$\{\phi_{ij}\}$. A periodicity 
in $\{\phi_{ij}\}$ implies a 
periodicity in the
physically measurable 
Josephson currents
$(\sin \phi_{ij})$.
It is easy to determine, analytically, 
the ground state of small fragments of the 
lattice. As before, we enforce the 
constraints via Lagrange multipliers;
now we leave the extrema equations
in their plain form and resort to solving 
a few trigonometric equations and, in the aftermath,
compare contenders amongst the resulting constrained extrema (along with
various states that lie on the boundaries of 
$\{\Pi_{bonds} \otimes \phi_{ij}\}$
of the different topological charge sectors:
scenarios in which a few
of the bond angles are $\phi_{ij} = \pm \pi$). 

\begin{figure}[htbp]
 \begin{center}
     \epsfxsize0.4\columnwidth
\epsfbox{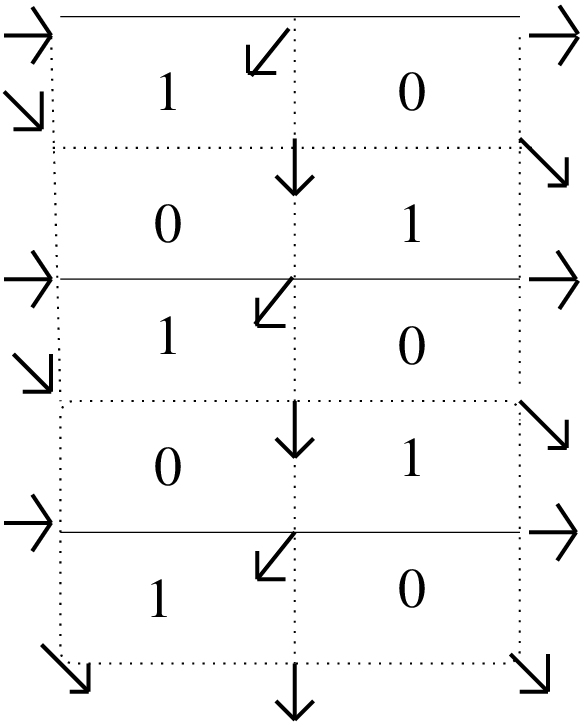}
\caption{A ground state of
the Fully Frustrated XY model 
in an explicit gauge. The arrows denote
the rotor directions. The numbers (1 or 0) at
the centers
of the plaquettes denote 
their topological charge.
The solid lines denote antiferromagnetic 
bonds and the dashed lines 
denote ferromagnetic bonds.
Any uniform global rotation
will produce another
ground state as 
will an interchange
between
sites on the 
two  
sublattices- a 
translation by
one lattice
constant.}  
\label{fig:explicit}
\end{center}
\end{figure}

Let us start by considering a 
single plaquette with open boundary conditions. 
If this plaquette is unfrustrated, the ground state
is trivially a ferromagnet: all $\phi_{ij} =0$
(this state lies in the $m=0$ sector). 
If this plaquette is threaded by half a
fluxon ($f=1/2$) then 
within the ground state the gauge
invariant angles are all the same
$\phi_{12}=\phi_{23}=\phi_{34}=\phi_{41}=\pm \pi/4$
where $i=1,2,3,$ and $4$ are the vertices
of the plaquette.
(Note that the directions in these bond angles
are important $\phi_{ij} = -\phi_{ji}$).
If all angles are $\pi/4$ then the plaquette 
has topological charge $q=1/2$ or a vorticity
$m=1$; if all
bond angles are $-\pi/4$ then the plaquette
charge is $q=-1/2$ and the
lies in the $m=0$ sector. 
Other topological charge sectors
are elevated relative to the
ground state by a finite gap.
For instance, the minimum energy
for the unfrustrated plaquette ($f=0$)
subject to the constraint
$m=1$ is $\epsilon_{min}^{m=1}=-1/2$ 
(with a configuration having
$\phi_{ij} = \pi/3$ for three bonds
and with the remaining bond on the 
boundary of the angular zone 
($\phi_{ij} = \pi$)) lies
above the ground state $\epsilon_{min}^{m=0} =-4$. 
Thus, each state in the $m=1$ sector is elevated
by at least $7/2$(J- if units are restored) relative to the ground
state\cite{lat}. Similar bounds may be generated in frustrated
plaquettes. These finite energy gaps will allow us, later
on, to argue for finite 
temperature vortex ordering 
for certain values of $f$ 
and to prove the existence of 
a $Z_{2}$ phase transition 
for the Fully Frustrated XY model
observed numerically in \cite{Olsson}. 
Note that here the ground states
computed with open boundary conditions
can trivially tile the plane.
One of the complications of the
continuous XY spectrum- {\it unlike
that of the approximate Coulomb charge dual}-
is that the various topological
charge configurations partially overlap and are {\it nested}.
The energies of the single unfrustrated 
plaquette span the region 
$ -4 \le \epsilon(m=0) \le 4$
when $m=0$ and 
$-1/2 \le \epsilon(|m|=1) \le 4$
when a vortex/antivortex is present.
The energy of a vortex free plaquette
may be higher than that of a 
vortex-full plaquette! The {\em nesting} of 
the energy spectrum just found hints that the 
XY model cannot be probed
by merely replacing its
plaquettes by topological charges.
This subtle non equivalence 
between the continuous model 
and its discrete counterpart
is not heavily emphasized
in the literature. The dual model (derived 
by various approximations for the 
{\em low energy sector}
in Eqs.(\ref{topo1},\ref{topo3}))
is a useful tool but it is merely
an approximation albeit an amazingly
beautiful and powerful one. 
The complication arising from the non-equivalence
of these models will be one 
of major obstacles
that we will need to 
surmount. As noted, the only ground states for a uniform $f=1/2$ 
are the two checkerboard patterns where 
$\phi_{ij \in R} = \pm \pi/4$
for all plaquettes $R$. The proof 
is quite simple: the energy
\begin{eqnarray}
E = \frac{1}{2} \sum_{\alpha} \epsilon_{\alpha} \ge \frac{1}{2}
\sum_{\alpha} \min \{ \epsilon_{\alpha}\},
\end{eqnarray}
where $\epsilon_{\alpha}$ is 
the energy of each individual plaquette.
Thus if a configuration in 
which for each plaquette
$\epsilon_{\alpha} = \epsilon_{\min}$ 
exists then that configuration is 
a minimum energy configuration. 
For each individual plaquette $R_{\alpha}$,
if $\epsilon_{\alpha} = \epsilon_{min}$
then all $\phi_{ij \in \alpha}= \pm \pi/4$.
Now, a bond $ \langle ij \rangle$ is common to two adjacent 
plaquettes. If its value, as read counterclockwise
from the center of one plaquette, is $\pi/4$ then
its value as read counterclockwise from the 
center of the adjacent plaquette is $-\pi/4$.
If we start from one plaquette
in which all bond angles are $\pi/4$
and then add an adjacent plaquette and 
require that the in the new additional plaquette 
all bond angles will be of equal
value which is set to either $\pi/4$ or
$-\pi/4$ then we will trivially see
the bond common to both plaquettes
will force all bonds in the new plaquette
to be equal to $-\pi/4$. We can then 
add a new additional plaquette to the last one and
repeat the argument. Thus if within the ``seed plaquette''
(the plaquette situated at the origin of the dual 
lattice) all gauge invariant bond angles 
$\phi_{\langle ij \rangle}= \pi/4$ 
then any other plaquette lying at a point $\vec{x} = (x_{1},x_{2})$
will also saturate the minimum energy bound
with all bond angles set to $[(-1)^{(x_{1}+x_{2})}] \pi/4$. 
We could have similarly started with a 
plaquette at the origin in which all bond angles
were $-\pi/4$. These are the only two 
possible ``seeds'' and the resultant configurations
are the only two global ground states. Within these 
two ground states we find a checkerboard 
pattern of topological charges:
If, in a given plaquette $R$, all the 
gauge invariant bond angles $\phi_{ij} = \pi/4$
then the vorticity $m_{R}=1$; when all $\phi_{ij} = -\pi/4$
there is no vorticity ($m_{R}=0$). The one
and zero values of the local vorticities form 
a checkerboard matrix $m_{\vec{x}}= \pm \frac{1}{2}(1+(-1)^{x_{1}+x_{2}})$.
This phase is a specific example of a staggered 
flux phase investigated in the High Temperature 
Superconductivity literature \cite{HTSC}.
An explicit ground state configuration
is shown in Fig.(\ref{fig:explicit}).

To distinguish between the two 
viable ground states and to highlight 
their $Z_{2}$ symmetry and pinpoint
the breaking of this symmetry a 
staggered magnetization  
\begin{eqnarray}
M \equiv | \sum_{\vec{x}} (-1)^{x_{1} + x_{2}}  m_{\vec{x}} |
\end{eqnarray} 
is defined \cite{Olsson}.
The current patterns within
the two different chiral ground states
are rotated 
by $\pi/2$ relative to
each other.
Domain walls separate the two checkerboard ground states- 
these cannot be transformed into each other by a continuous 
rotation. The walls consist of links which separate
two plaquettes with the same sign 
of the chirality. 
In some numerical studies 
the onset of XY symmetry 
``breaking'' is extracted 
from an analysis of the 
helicity modulus. As mentioned earlier,
the helicity modulus exhibits a
universal jump at the Kosterlitz-Thouless
transition \cite{He}.
The helicity modulus may be 
measured by applying 
strains on the Josephson 
currents \cite{Straley}. 
Let us next consider
a $2 \times 2$ block
with $f=1/2$ everywhere.
Such a small fragment
is depicted in 
Fig.(3)
where, in the chosen gauge,
$A_{25}=A_{58}=\pi$
with all other $A_{ij}=0$.

\begin{figure}[htbp]
\begin{center}
 \epsfxsize0.3\columnwidth
\epsfbox{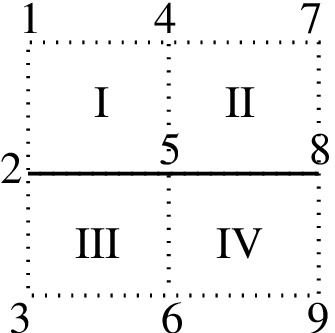}
\label{fig:block}
\caption{A 2 $\times 2$ block with
$f=1/2$: $A_{25}=A_{58}=\pi$.}
\end{center}
\end{figure}

Within the open boundary condition minimum of this 
small $2 \times 2$ block:
$\phi_{21}=\phi_{14}=\phi_{74}=\phi_{87}=
\phi_{89}=\phi_{96}=\phi_{36}=\phi_{23}= 
\cos^{-1}(\frac{2}{\sqrt{5}})$ and 
$\phi_{45}=\phi_{52}=\phi_{65}=\phi_{58}=
\cos^{-1}(\frac{1}{\sqrt{5}})$.
This state lies in the topological sector 
$(m_{I}=0,m_{II}=1,m_{III}=1,m_{IV}=0)$.
A similar state appears in the sector $(1,0,1,0)$.
The energy of these states is 
$\epsilon_{0,1,0,1}^{\min} = \epsilon_{1,0,1,0}^{\min}=-4
\sqrt{5}$. The energy per bond in this
state is lower than that of the global 
(entire lattice) ground state minimum.
This is hardly surprising: open boundary
conditions were placed on this
small $2 \times 2$ block. At 
worst the energy of this block 
would have coincided 
with the global ground 
state energy.

Let us denote the amount 
by which the open 
boundary condition 
state improves on
the global ground
state configuration
(as placed on this
small $2 \times 2$ 
block) by 
\begin{eqnarray}
\Delta_{-} = 4 \sqrt{5} - 6 \sqrt{2}  = 0.458991...
\end{eqnarray}
The minimum of all $(1,1,0,0)$ or $(0,0,1,1)$
states is $\epsilon_{1,1,0,0}^{\min} = \epsilon_{0,0,1,1}^{\min}  =
-8$
(corresponding to $\phi_{25}= \phi_{58} =\pi$ with all
other gauge invariant bond angles $\phi_{\langle i j \rangle} =0$).
The lowest energies of all other non checkerboard 
topological sectors are even 
more elevated than those of 
the $(1,1,0,0)$ or $(0,0,1,1)$
sectors. Thus the energy of all
non-checkerboard configurations
are elevated by at least 
\begin{eqnarray}
\Delta_{+} = 6 \sqrt{2} - 8 = 0.485281...
\label{delta+}
\end{eqnarray}
relative to the global 
ground state
configuration.

\section{$Z_{2}$ Symmetry Breaking in the Fully Frustrated Model}
\label{FFXY}

We described the ground states
of the Fully Frustrated XY model
on the square lattice. The topological charges
behaved as they should- repelling each other
and in the process generating a 
checkerboard pattern.
We now prove that the uniformly 
frustrated XY model has a
$Z_{2}$ phase transition: a claim that, 
till now, has only
been verified numerically and 
suggested by approximate
schemes \cite{Olsson}. 
We show that the existence of 
a finite temperature phase transition 
is a consequence of finite energy domain 
walls between the two 
ground state domains. Very interesting 
earlier works examining topological
excitations (including domain walls)
are found in \cite{hal,Granato}.

\subsection{Energy Gaps on the Square Lattice}
\label{gaps}

The topological charge sectors' minima 
are separated from each other by 
finite energy gaps. For the $2 \times 2$
block in (Fig.~\ref{fig:block}) 
the lowest energy configuration which
does not belong to the ``good'' sectors
$\{m_{I}=1,m_{II}=0,m_{III}=1,m_{IV}=0\}$
or $\{m_{I}=0,m_{II}=1,m_{III}=0,m_{IV}=1\}$ 
topological charge sectors is given by 
$\phi_{14}=\phi_{47}=\phi_{78}=\phi_{89}
=\phi_{96}=\phi_{63}=\phi_{32}=0$
and $\phi_{25}=\phi_{58}= \pi$ and 
it lies in $\{ m_{I} =0,m_{II}=0,
m_{III}=1,m_{IV}=1 \}$. The energy of this 
state is $E_{0,0,1,1}^{min}= -8$. Similarly if
$\phi_{25}=\phi_{58} = \epsilon -\pi$ (with $\epsilon= 0^{+}$)
a state of a similar energy lies in 
$\{ m_{I}=1,m_{II}=1,m_{III}=0,m_{IV}=0 \}$.
$E_{1,1,0,0}^{inf} = -8$. Thus the infimum of
the energy in all other ``bad'' topological 
charge sectors is $\epsilon^{inf}_{bad} = -8$.
The $\{ \phi_{ij} = \pm \pi/4 \}$ states, although they 
are the global minima
configurations, are not the lowest energy configuration 
on this fragment of the lattice with open boundary conditions. 
Nonetheless, these {\em{reference}} states lie in the ``good sectors'' 
$\{m_{I}=0,m_{II}=1,m_{III}=0,m_{IV}=1\}$ 
and the $\{ m_{I}=1,m_{II}=0,m_{III}=1,m_{IV}=0 \}$.
The energy of this state on a $2 \times 2$ block
is $\epsilon^{ref}_{1,0,1,0}=-6 \sqrt{2}$. 
In a system with periodic boundary conditions
\begin{eqnarray}
E = \frac{1}{L_{c}(L_{c}+1)} \sum_{Blocks} \epsilon_{\lambda}
\end{eqnarray} 
where $\epsilon_{\lambda}$ is the net energy of all bonds lying in an 
$L_{c} \times L_{c}$ block $\lambda$. The sum is performed over 
all overlapping
$L_{c} \times L_{c}$ blocks. The multiplicative factor in
front stems from the fact that each bond will be included in the 
sum $L_{c}(L_{c}+1)$ times if one were to cover a square lattice,
with periodic boundary conditions, with all overlapping 
$L_{c} \times L_{c}$ blocks.  We foucus on the simplest case: $L_{c} = 2$. 
In Eq.(\ref{delta+}), $\Delta_{+} 
\equiv \epsilon^{inf}_{bad} - \epsilon^{ref}_{1,0,1,0}$,
with $\epsilon^{ref}$ the energy of a
{\em reference} global ground state 
configuration (i.e. $\{\phi_{ij} = \pm \pi/4\}$)
when evaluated on a finite $L_{c} \times L_{c}$ plaquette block.
Within each small block $\lambda$, for states of the  ``bad'' incorrect 
topological charge
registry:
\begin{eqnarray}
\Delta_{+} \le \epsilon^{\lambda}[\{m_{x_{1}x_{2}}\} \neq \frac{1}{2}
(1+ (-1)^{x_{1}+x_{2}})~~ \nonumber
\\ \mbox{or}~~\frac{1}{2}(1+(-1)^{x_{1}+x_{2}+1})] - \epsilon^{ref}
\end{eqnarray}
where, as before, $(x_{1},x_{2})$ label the plaquette coordinates 
in the plane.
Averaging the last set of inequalities over the entire
lattice ($\Lambda$):
\begin{equation}
E^{\Lambda}[\{m_{x_{1} x_{2}}\}] \ge E_{0}^{\Lambda}+
\frac{\Delta_{+}}{L_{c}(L_{c}+1)} N_{b} -
\frac{\Delta_{-}}{L_{c}(L_{c}+1)} N_{good}
\label{impor}
\end{equation}
where $E_{0}^{\Lambda}$ is the global ground state energy 
(of the lattice with 
periodic boundary conditions), $N_{b}$ denotes the number 
of blocks with incorrect $\{m_{x_{1} x_{2}}\}$ registry,
while $N_{good}$ denotes the number of 2$\times $ 2 
blocks with the correct checkerboard 
registry. Henceforth, we 
denote 
\begin{eqnarray}
\Delta \equiv \Delta_{+} - \Delta_{-} > 0.
\end{eqnarray}
Similarly, for the unfrustrated ($f=0$)
system, an $|m|=1$ plaquette ($L_{c}=1)$ with open b.c. 
is elevated by, at least, $\Delta = \frac{7}{2}$ 
relative to the $m=0$ ground state 
configuration. Here we employ the last
relation to obtain a lower
bound on the energy gap 
given the number of 
blocks (plaquettes in the $f=0$ case)
which do not have the 
correct topological charge
registry (i.e. have nonzero vorticity $m \neq 0$). 
The reader can see
how these bounds may be generalized to 
higher order commensurate values of 
$f$ where the block size 
$L_{c}$ will be larger. 
These simple inequalities
can be seen as a generalization
of the Peierls bound \cite{Prls}. 

We are now ready to apply the machinery of
reflection positivity. Let us quickly outline
the underlying logic: We assume that
at infinity we have  ``good'' $g_{2}$ boundary
conditions, by which we mean that 
one of the two types of topological 
charge checkerboard configurations
will be found. We consequently bound the 
size of the region that is 
occupied by the other ``good'' checkerboard 
ground state $g_{1}$.
In essence, we bound the size of the 
region separating the two checkerboard
configurations; in this region the 
topological charge configurations
are ``bad'' and costly in energy.
The number $N_{b}$ of the blocks
which have ``bad'' topological charge
configurations is, at least, linear in the 
perimeter of the regions ({\em domain walls}) 
surrounding the $g_{1}$ droplets. The correspondence
to the Peierls proof of 
a phase transition in 
an Ising ferromagnet 
becomes transparent 
toward the end of the 
proof. We avoid Peierls' contour inversion 
\cite{Prls},
opting instead for a direct
computation of the 
ratio of partition
functions in order 
to find the 
probabilities
of various 
contours.
We prove, trivially, that 
at sufficiently low temperatures
the energy bounds on the cost of 
``bad'' regions (domain walls) 
translate into bounds on their 
probabilities.

\subsection{Reflection Positivity}
\label{RP}

We employ Reflection Positivity \cite{RP}
as applied to a system invariant
under a reflection about an 
axis $P$ passing through 
lattice sites. The uninitiated reader is referred to Appendix A
for a brief exposure to this technique.
We partition the plane into two sides 
which flank the axis $P$ (which itself 
lies along a line of bonds):
$P_{\pm}$. Let us define ${\cal{P}}_{\pm}$
as the union of $P_{\pm}$ with $P$.
Under reflection about $P$:
\begin{eqnarray}
R_{p} F(\vec{S}(\vec{x}_{1}),...,\vec{S}(\vec{x}_{n})) = 
F(\vec{S}(R_{p}\vec{x}_{1}),...,\vec{S}(R_{p}\vec{x}_{n}))
\end{eqnarray}
with $\vec{x}_{i}$ planar locations.
Thus, under reflection 
\begin{eqnarray}
R_{p}A_{ij} = A_{Ri,Rj}, ~R_{p}\theta_{i}=\theta_{Ri} \nonumber
\\ \mbox{if $\langle ij \rangle$ is vertical:}
~R_{p}\phi_{ij}= \phi_{ij} \nonumber
\\ \mbox{if $ \langle ij \rangle$ is horizontal:}
~R_{p} \phi_{ij} = -\phi_{ij}.
\end{eqnarray}
For both horizontal and vertical bonds, $\cos \phi_{ij}$ is invariant
under $R_{p}$. Defining 
\begin{eqnarray}
B \equiv -[\sum_{\langle ij \rangle 
\in {\cal{P}}_{\pm}} -\frac{1}{2}\sum_{\langle
  ij \rangle \in P}] \cos \phi_{ij},
\label{rp_defn}
\end{eqnarray}
the Hamiltonian
$H = B+ R_{p}B$ is manifestly invariant under $R_{p}$ 
(yet the reflected plaquettes correspond to the same 
frustration $0 \le f \le 1/2$ only when $f=0,1/2$). 
As the sense of the circulation in the sum $\sum_{\Box R} \phi_{ij}$
is fixed over the entire lattice, under $R_{p}$:
$(m_{\vec{x}}-f) \rightarrow -(m_{\vec{x}}-f)$.
For $f=\frac{1}{2}$, the vorticities transform as
\begin{eqnarray}
m=-1,0,1,2 \rightarrow 2,1,0,-1
\end{eqnarray}
under reflections. The correct (ground state pattern) set of
topological charges $\{ m_{\vec{x}} \}$ 
at the various planar locations $\vec{x}$  
is invariant under $R_{p}$. This is to be expected as the energies 
are unchanged by $R_{p}$- there is no energy 
increase along the reflection axis $P$. When $f=0$, the vorticities
transform as
\begin{eqnarray}
m= -1,0,1 \rightarrow 1,0,-1
\end{eqnarray}
under $R_{p}$. 
For $0<f<1/2$, the quantity $2 \pi[-(m-f)]$ is not an allowed value for the circulation sum.
Let us define $b_{k}$ as the $L_{c} \times L_{c}$ topological charge pattern
about the central plaquette at $\vec{x}_{k}$, and introduce a function
$\omega_{b_{k}}(\vec{x})$ which is equal to one if $b_{k}$ is found 
within the $2 \times 2$ block centered 
about $\vec{x}$, and is equal to zero otherwise.
Let us next reflect $b_{k}$ to cover the entire lattice 
$\Lambda$ and generate a vortex pattern $C_{k}$. One 
may define a partition function ($Z_{C_{k}}$)
that is the sum of all of the Boltzmann weights corresponding to
these configurations. There are two correct 
(checkerboard) topological charge configurations
within the $L_{c} \times L_{c}$ block; we shall denote these
topological charge configuration by $g_{1}$ and $g_{2}$.
For all faulty (non $g_{1}$ or $g_{2}$) blocks $b_{k}$ 
reflections generate $N_{b} >t^{\prime}N$ 
faulty $L_{c} \times L_{c}$ blocks within the
lattice with $t^{\prime}>0$.
As before, let $\{ \lambda_{i}\} \equiv$ 
vortex~patterns~in~the $L_{c}
\times L_{c}~$ blocks centered about the sites $i$,
and let $\{g_{1},g_{2}\}$ denote the two correct (checkerboard)
vortex configuration within an $L_{c} \times L_{c}$ block. 
Note that the set $\{ \lambda_{i}\}$ consists
of all {\em{maximally overlapping}} $L_{c} \times L_{c}$ blocks.

\subsection{The Kaleidoscope Patterns: Energy Gaps on The Reflected Blocks}
\label{kld}

By repeatedly reflecting any $L_{c} \times L_{c}$ 
block $\lambda$ to cover
the entire lattice we
generate a periodic pattern
of repetetive $2L_{c} \times 2L_{c}$ super-blocks. 
The planes $P$ may be
either vertical or horizontal and for 
each orientation $R_{p~horizontal}^{2} = R_{p~vertical}^{2} = 1$.
After two reflections
by any two parallel planes the 
original pattern must emerge. 
Taking note of all possible 
``bad'' (non-checkerboard) $L_{c} \times  L_{c}$
blocks one sees that in all
cases the global pattern $C_{k}$ generated by
consecutive reflections contains
more bad $L_{c} \times L_{c}$ overlapping blocks
than those of the checkerboard type:
in Eq.(\ref{impor})  $N_{b} > N_{good}$.
Consequently, the energies
of the all repetitive patterns
satisfy
\begin{eqnarray}
E_{C_{k}} > E_{0}^{\Lambda}+
\frac{\Delta}{L_{c}(L_{c}+1)} 
N_{b} \nonumber
\\ >  E_{0}^{\Lambda} + \frac{\Delta}{L_{c}(L_{c}+1)} 
t^{\prime} N,
\end{eqnarray}
with $L_{c}=2$, and, as
denoted earlier, with 
the number of bad blocks
in each repetitive pattern
$N_{b} > t^{\prime} N$.

\subsection{Chessboard Estimates}
\label{chess}

Let us next define an angular volume $\Omega_{\lambda}$ 
within a $ L_{c} \times L_{c}$ block such that
$\sup_{\theta \in \Omega_{\lambda}}
|\theta_{i}-\theta^{ ground }_{i}| = \delta$
for all sites $i$
(where $\theta^{ ground }_{i}$ denotes
the angular configuration of
a global ground state minimum on the 
entire lattice-
the pinned down reference angular configuration
with broken global $Z_{2}$ and $U(1)$ symmetries)
and, correspondingly, 
\begin{equation}
\max_{L_{c} \times L_{c}  \mbox{block partitions}~ \lambda} 
\sup\{\epsilon_{\lambda}\} =
\epsilon_{ref} + \epsilon_{0}.
\end{equation} 
The size of $\Omega_{\lambda}$ is $|\Omega_{\lambda}| = (2 \delta)^{9}$. 
The volume $\Omega_{\lambda}$ is strictly contained 
in a single checkerboard sector. On the fringes ($\partial \Omega_{\lambda}$)
of this volume:
\begin{equation}
\epsilon|_{\partial \Omega_{\lambda}} \le \epsilon^{ref}
+ \epsilon_{0},
\end{equation}
with $\epsilon^{ref}$ the energy of a
global ground state configuration 
when evaluated on a finite $L_{c} \times L_{c}$ plaquette block,
and $\epsilon_{0} < \Delta$.
Next, we define a global angular
volume $\Omega$ such that 
$\sup_{\theta \in \Omega}
|\theta_{i}-\theta^{ ground }_{i}| = \delta$
for all lattice sites $i$.
The energy of any global lattice
configuration within this 
volume has an energy which is bounded
from above by 
\begin{equation}
E_{\partial \Omega} \le E_{0}^{\Lambda} + \frac{N
\epsilon_{0}}{L_{c}(L_{c}+1)}.
\end{equation} 
The definition of $\Omega$ and $\Omega_{\lambda}$ 
is made possible by the continuity in $\{\theta_{i}\}$ of 
the Hamiltonian in Eq.(\ref{definition}).
Now 
$[Z_{C_{k}}/Z_{tot}]$
may be easily bounded at low temperatures
\begin{equation}
\frac{Z_{C_{k}}}{Z_{tot}} < \frac{Z_{C_{k}}}{Z_{\Omega}} < 
\frac{(2 \pi)^{N}}{(2\delta)^{N}} \exp \Big[-\beta t^{\prime} N
\frac{\Delta- 
\epsilon_{0}}{L_{c}(L_{c}+1)}\Big]
\end{equation}
where $(2 \pi)^{N}$ is the net angular volume
of the entire system and $(2 \delta)^{N}$ 
is the volume of  $\Omega$. Here we 
noted the angular volume of 
$C_{k}$ is bounded by $(2 \pi)^{N}$. \cite{sophisticated}  

Let us define a ``logical'' test function 
on a block $\lambda_{i}$ centered 
about site $i$ by 
$F(\lambda_{i}=g_{1},g_{2})=1$ and 
$F(\lambda_{i} \neq g_{1,2})=0$  
(the reader will note that is merely a variant
of $\omega_{b_{k}}(\vec{x})$: $F =  \omega_{g_{1}} + \omega_{g_{2}}$).
Let us furthermore define
\begin{eqnarray}
V(\vec{x}) \equiv \prod_{\vec{x} \in \lambda_{i}} F(\lambda_{i}).
\end{eqnarray}
The region where $V(\vec{x})$ vanishes defines a domain 
$D$. In the spirit
of providing very generous upper
bounds, let $\ell$ denote the outer perimeter of $D$, and let us 
pick merely
\begin{eqnarray}
\ell^{\prime} \equiv \max\{\mbox{Int} [\frac{\sqrt{\ell}}{8 L_{c}}],1\}, 
\end{eqnarray}
non-overlapping faulty $b_{k}$ 
(where Int$[~]$ denotes
the integer part). The probability of a domain $D$,
\begin{eqnarray}
Prob(D) \le \max \bigl \langle 
\omega_{b_{1}}(\vec{x}_{1})...\omega_{b_{\ell}}(\vec{x}_{\ell}) \bigr 
\rangle \nonumber
\\ \le \prod_{k=1}^{\ell^{\prime}} [Z_{C_{k}}/Z_{tot}]^{W/N}  <
\exp[-\beta W t \ell^{\prime}]
\label{bound}
\end{eqnarray}
with
\begin{eqnarray}
t \equiv [t^{\prime} (\Delta- \epsilon_{0}) - \beta^{-1}
 \ln(\pi/\delta)]/(L_{c}(L_{c}+1)).
\label{change}
\end{eqnarray} 
In the last inequality, the blocks $\{\lambda_{\ell}\}$ are
chosen to be {\em non-overlapping}: no spins 
are shared by two blocks (including the boundaries 
of the blocks). The second inequality in 
Eq.(\ref{bound}) followed from reflection 
positivity. (The reader is referred to 
equation (\ref{multi})
in Appendix A for details.) Let $G^{i}$ ($i=1,2$) denote the support
of the correct vortex configuration ($g_{i}$) in a region of any size. Let us
assume that at infinity, we have $G^{2}$ boundary conditions.
The probability of having a domain $D$ incurred
by $N_{G^{1}}$ units (in a partitioning of the 
lattice to {\em{non overlapping}} blocks) of the opposite checkerboard 
state is 
bounded by Eq.(\ref{bound}). The bounds
on the area covered  
by the opposite checkerboard state
may proceed as in the usual Peierls 
argument (see Appendix B  and Eq.(\ref{standard}) for details),
to yield 
\begin{eqnarray}
r \equiv \langle \frac{N_{G^{1}}}{N} \rangle <  
\sum_{\ell = 4 L_{c},...} \ell^{2} 3^{\ell} \nonumber
\\ \times \exp[-\beta Wt ~\max\{\mbox{Int}[\frac{\sqrt{\ell}}{8 L_{c}}],1\}].
\label{Peierls}
\end{eqnarray} 
The right hand side can made as small as desired for 
$\beta >\beta_{c}(r).$
Stronger bounds hold 
for all other ``bad'' topological
charge configurations. Thus,
at sufficiently low
temperatures, a spontaneous 
symmetry breaking is sparked 
by applying boundary 
conditions
at 
infinity. 

Analogously, for $f=0$, the fraction of $|m|=1$ vortices 
(which, for $L_{c}=1$, are elevated by at least
$\Delta = 7/2$ relative to the lowest lying 
$m=0$ configuration)is bounded from 
above by arbitrary, exponentially small, numbers at 
sufficiently low temperatures: ``pair annihilation''.

Here we avoid the contour 
flipping algorithm of the 
standard Peierls argument
by use of Reflection 
Positivity to bound 
$Prob(D)$ in Eq.(\ref{bound}). 
We kept $L_{c}$ general
(and have not explicitly set
$L_{c}=2$) in order 
to suggest other plausible extensions 
(not hinging on Reflection Positivity) 
to other values of $f$ 
for which $L_{c}$ will 
be larger). To emphasize-
in this proof a necessary (but not sufficient
requisite) is to 
establish a gap between
the minimum energy per site 
of ``bad'' topological 
configurations on the $L_{c} \times L_{c}$ 
block with open boundary conditions
and thermodynamic limit ground 
state tiling the entire 
plane belonging to the ``good'' 
topological sector (which 
we dubbed the ``reference'' state). 
In general, the open boundary 
condition minimum on $\lambda$
will cannot tile the entire 
plane and will produce a slightly 
lower ground state energy per site 
than the ground state of the 
entire lattice.

\section{An Alternative Proof}
\label{altern}

The reader might be a bit
dismayed that our 
proof seems to rest on 
the relatively fortuitous
event that 
for the small $L_{c}=2$ the gap $\Delta \equiv \Delta_{+} - \Delta_{-}>0$.
Had this not been the case we might have
been required to compute 
energies on much larger 
$L_{c} \times L_{c}$ 
blocks. Moreover, in the previous 
proof, in subsection(\ref{kld}),
an explicit 
check that the number
of bad blocks $N_{b}$
is greater than the 
number of good blocks 
was necessary for the 
configurations generated
by consecutive reflections
of $\{b_{k} \}$ to
cover the entire lattice.
Here we show that this is 
not the case when attacking
the problem along a slightly 
disparate
path. In the spirit of Eqs.(\ref{rp_defn},\ref{rp-symm}) let us define 
\begin{eqnarray}
\epsilon_{\lambda^{\prime}} \equiv  \sum_{\langle ij \rangle \in R} \cos \phi_{ij}
-\frac{1}{2}  \sum_{\langle ij \rangle \in  \partial \lambda^{\prime}}
\cos \phi_{ij}
\label{marginal}
\end{eqnarray}
for each $s \times s$ block $\lambda^{\prime}$ 
(with an arbitrary integer $s>1$) such that 
\begin{equation}
E = \sum_{\lambda^{\prime}} \epsilon_{\lambda^{\prime}}
\end{equation}
in a partition of the lattice 
with ``marginally'' non overlapping blocks $\lambda^{\prime}$
(nearest neighbor blocks share only 
a common boundary $\partial \lambda^{\prime}$). 
The form Eq.(\ref{marginal}) is particularly suited
for Reflection Positivity treatments.
Note that the open boundary condition 
ground state $\{ \theta_{i} \}$ of $\epsilon_{\lambda^{\prime}}$ 
on the  block $\lambda^{\prime}$ 
(the previous ``reference state'')
may be repeatedly reflected (as $f=1/2$ is congruent to $(-f)$)
along the boundaries $\partial \lambda^{\prime}$
to tile the entire plane. Thus the open boundary
condition minimum on the 
small block is, in this
case, a portion of 
the global ground 
state.

Now, a finite energy gap $\Delta>0$ 
(necessitated by our definition of ``bad'' 
and ``good'' topological sectors) separates 
the various ``bad'' topological charge 
configurations (which do not 
contain a ground state of  
$\epsilon_{\lambda^{\prime}}$ on
the finite $s \times s$ slab
with open boundary conditions)
and the ground state value of 
$\epsilon_{\lambda^{\prime}}$. 
The reader might recognize 
$\Delta$ as none other
than $\Delta_{+}$ for
the particular case 
$s=L_{c}$ as defined 
for the energy in 
Eq.(\ref{marginal}).
In computing the probability
of a domain $D$ we note 
that net domain size 
$N_{b}^{\prime}$ depends
on how we partition the
plane into marginally
non overlapping blocks $\lambda^{\prime}$. 
If, in a partitioning
of a lattice into 
maximally overlapping
blocks $\{ \lambda \}$,
we find $N_{b}$ faulty
blocks of the incorrect
topological charge registry
then in the optimal
partitioning of the 
lattice into marginally
non overlapping blocks 
$\{ \lambda^{\prime} \}$
we will find, at least, 
\begin{eqnarray}
N_{b}^{\prime} = \mbox{Int} \Big[ \frac{N_{b}+s^{2}-1}{s^{2}} \Big]
\label{least}
\end{eqnarray}
faulty blocks $\{\lambda^{\prime}_{b}\}$.
Consequently, in Eq.(\ref{change}),
we may substitute 
\begin{eqnarray}
\Delta \to \frac{\Delta}{s^{2}}
\label{substitute}
\end{eqnarray}
and repeat the previous
arguments of section
(\ref{FFXY}).

Although, in the 
above, we
focused attention on
$s \times s$ 
square blocks, we could
have chosen 
$\lambda^{\prime}$
to be a domino:
the minimal 
$(2 \times 1)$ plaquette
configuration. There are
two checkerboard $g_{1,2}$
states (the 
sectors $(m_{1},m_{2}) = (1,0)$ or $(0,1)$)
within the domino and
all of our arguments may be reproduced
with the factor $s^{2}$ of 
Eqs.(\ref{least},\ref{substitute})
replaced by the area 
of the domino (two plaquettes).

Thus, putting all of the pieces together, we have rigorously demonstrated, using various approaches, that at sufficiently low positive temperatures, the topological charges in the Fully Frustrated
XY model form arrangements which lift the $Z_{2}$ symmetry associated with the two-fold degeneracy associated with the two ground state vortex patterns. This rigorous result reaffirms earlier tour de force numerical and extremely insightful yet more approximate analytical treatments  \cite{teitel-review}.

\section{The Standard XY Model- Vortex Confinement}
\label{stand}

We may similarly bound the number of radius $r$ vortices ($r \gg 1$),
which were defined previously, within the 
unfrustrated system. When $r \gg 1$ the energy penalty is no longer 
a mere $7J/2$  as it was for a vortex of size $r=1$
(which was immediately adjacent to an antivortex),
but rather a forbidding logarithmic function of $r$. 
Employing the Reflection Positivity inequality Eq.(\ref{chessboard}) 
with the energy penalty associated with the
large radius vortex of Eq.(\ref{ar-vortex}), the probability of a 
vortex of size $r$
\begin{eqnarray}
Prob( r-vortex) ~ ~ < (r_{1}/r)^{\beta W K_{1}},
\end{eqnarray} 
where, we once again employed the 
fact that the volume of the state
generated by repeated reflections
of the radius $r$ vortex to cover 
the entire plane has an angular 
volume which is trivially
bounded by $(2 \pi)^{N}$.
This probability
can be made as small as desired at 
sufficiently low temperatures (large $\beta$). 
The probability
of a radius $r$ vortex drops algebraically
with $r$: There are no ``real'' macroscopic vortices
at sufficiently low temperatures. In \cite{clock}, 
rigorous results on the related low temperature Kosterlitz-Thouless phase
in $k>4$ state clock models (which, as we will further discuss
in Section \ref{Z_k}, become the XY model in the $k \to \infty$ limit) 
were arrived at by a combination of contour arguments, exact dualities,
and Griffiths' type inequalities.

\section{Kaleidoscope Field Patterns} 
\label{kaleid}

Although in many instances 
(e.g. uniform non commensurate 
$f$) the determination of the exact 
ground state is nontrivial
there are several  
exceptions to the rule. 

If the frustrations $\{f_{\lambda^{\prime}}\}$
form a kaleidoscope pattern-
i.e. if they may be generated, starting
from an arbitrary sequence of
frustrations in a small region, by consecutive
reflections in different
planes to cover the entire
lattice, then a ground state
may also be formed by the same 
sequence of reflections.
Let us now ``elaborate '' and prove
this statement. Let $\lambda_{1}^{\prime}$ 
be a block spanning one or
more plaquettes on which
the frustrations
at the different plaquettes
are arbitrary. If the 
block is of finite
size then the determination
of the exact open boundary
condition minimum 
of the energy 
functional $\epsilon_{\lambda_{1}^{\prime}}$
defined in Eq.(\ref{marginal})
will yield an angular
configuration(s) 
$\{ \theta_{i} \}$
on $\lambda_{1}^{\prime}$.  
Now let us invert
the frustrations and set
$f_{\lambda_{2}^{\prime}} =  -f_{\lambda_{1}^{\prime}}$
for the block $\lambda_{2}^{\prime}$ 
which is the mirror image 
of $\lambda_{1}^{\prime}$ about a
plane $P_{1}$ which 
passes through one of the 
edges of $\lambda_{1}^{\prime}:
\lambda_{2}^{\prime} = R_{P_{1}} \lambda_{1}^{\prime}$.
We now note that the open boundary 
condition minimum of $\epsilon_{\lambda^{\prime}}$
on the fused super-block $\lambda_{1}^{\prime} \cup
\lambda_{2}^{\prime}$
must satisfy
$\min\{ \epsilon_{\lambda_{1}^{\prime} \cup {\lambda_{2}^{\prime}}}\} 
= \min \{ \epsilon_{\lambda_{1}}+ \epsilon_{\lambda_{2}} \}
\ge [\min \{\epsilon_{\lambda_{1}}\} + \min \{\epsilon_{\lambda_{2}}\}]$,
yet just such a configuration saturating the 
lower bound with
an energy [$\min \{\epsilon_{\lambda_{1}}\} + \min
\{\epsilon_{\lambda_{2}}\}$]
may be constructed. To do so we may simply
leave the original angular configuration 
$\{\theta_{i}\}$ for all sites $i$ within the 
block $\lambda_{1}^{\prime}$
untouched and set $\theta_{i}  = \theta_{R_{P_{1}} i}$ for 
all sites $i$ in the reflected block $\lambda_{2}^{\prime}$.
We may now choose a different plane $P_{2}$
and keep repeating the process 
until $\lambda_{1}^{\prime} \cup \lambda_{2}^{\prime}
\cup \lambda_{3}^{\prime} \cup \ldots$ spans the entire lattice.
Perhaps the simplest realizations
of such ground states
are the staggered flux 
phases (of which 
the $f=1/2$ ground state
hitherto considered
is a special example).
As other realizations we may consider 
systems in which
the frustration  
\begin{eqnarray}
f_{\vec{x}} = {\mbox{f}}~ \sin(\frac{\pi x_{1}}{n})  \sin(\frac{\pi
x_{2}}{n})
\label{Periodic}
\end{eqnarray}
for all plaquettes $\vec{x}$ within the dual lattice
with an arbitrary f and a finite integer $n$.
Here $\lambda_{1}^{\prime}$ is the 
elementary $n \times n$ block. 
Any reflection
of the frustrations $f_{\vec{x} \in \lambda_{i}^{\prime}}$
inverts their sign 
on the sites $R_{p} \vec{x}$
which lie in $\lambda_{i+1}^{\prime}$. 
Many systems, such as that of Eq.(\ref{Periodic}),
are reflection positive
and a proof of a finite
temperature ordering 
similar to that for $f=1/2$
follows.

\section{Generalized $Z_{k}$ symmetry and Magnetic groups}
\label{Z_k}

For a uniform frustration $f= p/q$ with relatively prime $p,q \ge 3$,
the topological charge ground state consists 
of repetitive units 
extending $k_{1},k_{2}$ units along the $x_{1},x_{2}$ 
axis respectively. By translating 
the origin in the covering of the 
lattice by non overlapping blocks, we may generate $k= k_{1} k_{2}$ 
ground state configurations
within the repetitive blocks. To describe
this viable translational $Z_{k_{1}} \otimes Z_{k_{2}}$ 
symmetry, we 
may extend the definition
of the $Z_{2}$ order parameter 
previously defined: 
\begin{eqnarray}
M_{k} \equiv  |\sum_{\lambda} \sum_{g} \omega_{g}(\lambda)~ \exp[2 \pi i g/k]|,
\end{eqnarray}
where the sum is now over non overlapping
blocks $\lambda$ and possible 
ground state topological charge sectors
$g$. As before, $\omega_{g}(\lambda) = 1$ 
if the pattern $b_{g}$ appears in $\lambda$
and is zero otherwise. We may similarly define 
the two block correlator
$\langle \omega_{i}(\lambda) \omega_{j}(\lambda^{\prime}) \rangle$
for two blocks $\lambda$ and $\lambda^{\prime}$  appearing
in a partitioning of the lattice into non overlapping 
blocks. In the limit $q \rightarrow \infty$ 
(irrational $f$) the discrete 
translational symmetry transforms into a continuous symmetry 
\begin{eqnarray}
Z_{k_{i}}  \rightarrow_{k \rightarrow \infty} O(2),
\end{eqnarray}
and the translational symmetry group becomes  
the symmetry group of the two-torus.

Classical two-dimensional $Z_{k>4}$ clock models
exhibit two transitions 
at self-dual temperatures values \cite{clock}.
In the temperature regime between the higher and 
lower transition temperatures the system exhibits algebraic long range order with algebraically decaying correlations; below the lowest transition temperature, the system exhibit long range order while it is disordered at temperatures above the higher of the two transition temperatures.
In the large $k$ limit of the clock model, the lower transition temperature veers to zero and the algebraic Kosterlitz-Thouless phase appears all of the way down to zero temperature.

The  $Z_{2}$ and $O(2)$ symmetries present
in the $f=1/2$ XY model become two
$O(2)$ symmetries in the limit 
of large $q$ (continuous $f$). 
Physically, all possible translations 
of a given ground state configuration
amount to a rotation of the spins 
by an all possible 
angles. One might 
speculate that perhaps for 
irrational $f$, there is no
finite temperature vortex
ordering (no explicit $Z_{k_{i}} \to O(2)$
symmetry breaking) but
rather a finite temperature 
Kosterlitz-Thouless
like transition \cite{II}. 
Such a viable link between the degeneracy amongst states
generated by displacements and
those created by global
rotations is also found elsewhere.
In the Quantum Hall problem, 
the degenerate states may 
be labeled in terms of their 
angular momentum quantum numbers or, 
alternatively, by their magnetic 
translation operator
eigenvalues.
On a lattice (the 
Hofstader problem \cite{Hofstadter}) 
the correct degeneracy
is easily read off in 
terms of the magnetic 
translation operators;
the flux piercing the
fundamental cell is 
an integer multiple
of the elementary 
fluxon. An irrational flux $f$ might 
lead to an effective loss of commensurability 
effects (the size of the elementary cell
becomes of the order of the system size) 
and the equivalence between 
those degeneracies spawned 
by translations and those 
created by rotations may be restored.
The sole (albeit important) distinction
between the Quantum Hall problem
and the frustrated XY model is
that in the latter the magnitude 
of the wavefunction is pinned,
$|\psi(\vec{x})|=1$, whereas
in the Quantum Hall problem
both phase and amplitude 
vary.

In addition to the trivial $Z_{k_{i}}$ groups
generated  by displacements, one finds in 
certain instances a greatly enhanced symmetry
group. The ground states of the frustrations 
$f=1/3$ and $f=1/4$ are a case in point
\cite{teitel-review}. Here
a multitude of ``zero-energy domain walls''
appear. Phrased in our language this implies 
that nontrivial coverings of the lattice are possible
such that within all overlapping blocks $\lambda$ 
we find the reference global ground state minimum 
configuration. At low temperatures the infinite degeneracy is 
lifted by the free energy of spin waves.

\section{Isotropic to Nematic
Transitions}
\label{nematic}

Just as we were able to
bound, at low temperatures,
the number
of free vortices
within the unfrustrated
XY model we
may also prove that
free disclinations
in liquid crystal
system must condense
at sufficiently
low temperature;
such a confinement
of disclinations
occurs in the transition
between the isotropic and
nematic (or ``topologically'' ordered
\cite{toner}) phases. 
Similar arguments
apply for
other systems
having other 
order parameter 
symmetries.
Within the (unfrustrated) XY model, the $O(2)$ vortices are elevated
by a finite amount relative to the ground state. 
Similarly within lattice models
of liquid crystals the $RP^{2}$ 
topological defects (disclinations) 
of the headless director fields 
$n^{\! \! \! \! \! \leftrightarrow}$ must 
incur an energy cost relative to the 
defect free ground state.
This fused with a Reflection 
Positivity symmetry which 
is present in many of these models
(for some lattice and continuum models including those for which
of Reflection Positivity and other considerations were employed for proofs of intermediate phases
as evinced by angular correlation functions, see \cite{Lasher, nematic_RP_1,toner,nematic_RP_2,nematic_RP_3}),
allows us to set bounds via chessboard estimates
on the frequency of disclinations (the topological charges in nematic systems)
at  sufficiently low 
temperatures. 
Analogously,
one may examine
an analogue
of the Fully Frustrated 
XY model:
a nematic model 
on a lattice (as in \cite{toner}) having 
half an $RP^{2}$
fluxon thread each
of its plaquettes.
Similar arguments
may be used to bound
the frequency and size
of topological defects
associated with 
different systems.

\section{Conclusions}

We presented a theoretical
study of the XY model with specific emphasis
on the fully frustrated case. A main result
is a proof of the existence
of a $Z_{2}$ symmetry breaking transition 
within the fully frustrated XY model consistent
with previous numerical results and approximate analytical treatments available in the
literature \cite{teitel-review}. More generally, we discussed extensions
of the methods that we introduced here to 
other arenas and models where similar
binding and orders of topological charges appear (confinement
of disclinations in the isotropic
to nematic transition, topological charge
orders in various XY models and bounds on
vortices within the unfrustrated XY model). 
A key complication that our treatment overcomes 
is the non-exact equivalence between the energies of the continuous
angular system that we investigate and the energies of the system in different sectors of discrete topological charges; unlike the spectral character of the discrete dual topological charge system
with Coulomb type interactions, the energies of angular configurations
that lie in different topological charge sectors partially overlap with each other and 
are nested.

\acknowledgements

The author gratefully acknowledges illuminating
discussions with A. Auerbach, 
L. Chayes, and J. Zaanen.
Support by FOM and by
NSF CMMT 1106293 are gratefully
acknowledged.

\section{Appendix A: What is Reflection Positivity?}

Reflection Positivity has long been a tool
reserved for the cabinets and drawers of MPs (Mathematical 
Physicists). Our simple minded nuts and bolts 
approach presented below \cite{oldme} is geared 
toward our very specific limited set 
of problems and is quite superficial. 
It does the field no justice. For a 
more detailed and rigorous 
exposition the reader is referred to \cite{RP,marekp}.
We remark that in a more recent interesting work, Reflection Positivity has been employed 
to study quantum topological orders \cite{jaffe}.

To simply understand Reflection Positivity with a relative minimum
formalism, let us imagine that we partition our
model into two sides 
which flank an orthogonal bisecting plane $P$ 
(which itself lies along a line of bonds):
$P_{\pm}$. Let us define ${\cal{P}}_{\pm}$
as the union of $P_{\pm}$ with $P$.
In a Reflection Positive system, 
there exists an operator $B$ such that 
the Hamiltonian may be expressed 
as 
\begin{eqnarray}
H = B + R_{P} B.
\label{rp-symm}
\end{eqnarray}
As noted earlier in the text, the Reflection operator $R_{P}$
about the plane $P$ is defined by
\begin{eqnarray}
R_{p} F(\vec{S}(\vec{x}_{1}),...,\vec{S}(\vec{x}_{n})) \nonumber
\\ = 
F(\vec{S}(R_{p}\vec{x}_{1}),...,\vec{S}(R_{p}\vec{x}_{n})),
\label{Refl}
\end{eqnarray}
where $\vec{x}_{i}$ denotes the planar location of site $i$
\cite{construction}.

The basic inequality that we will need is given by the following statement: the thermal average  
\begin{equation}
\langle {\cal{O}} R_{p} {\cal{O}} \rangle \ge 0
\label{basic_ineq}
\end{equation}
for all functions $\{{\cal{O}}\}$. (In other
words, such thermal averages behave like 
inner products in a ``positive'' sense.) 

Proof: Let us write the 
thermal average
longhand 
\begin{eqnarray}
\langle {\cal{O}} R_{p} {\cal{O}} \rangle = Z^{-1} \sum_{\{\vec{S}_{i}\} \in
P_{+}} \sum_{\{\vec{S}_{i}\} \in
P_{-}} \sum_{\{\vec{S}_{i}\} \in
P}  {\cal{O}} R_{p} {\cal{O}} \nonumber
\\ ~ \exp[-\beta (B+ R_{p} B)], 
\label{positive}
\end{eqnarray}
where the partition function $Z>0$. Thus the sign of 
the thermal average is 
sign of the multiple sum which, by symmetry,
may be folded back onto one half plane to 
become a sum of squares when $\{{\cal{O}}\}$
is a local operator defined on ${\cal{P}}_{+}$
or on ${\cal{P}}_{-}$.
Let $F \equiv {\cal{O}} \exp[-\beta B]$
be an operator on ${\cal{P}}_{+}$. 
Let us explicitly segregate the spin arguments 
in ${\cal{P}}_{+}$ which  correspond to spins 
lying on $P$ (which we shall denote by $\{S(\vec{x}_{j})\}$)  and those 
that lie in $P_{+}$ (labeled by $\{S(\vec{x}_{i})\}$). Note 
that (Eq.(\ref{Refl})) $R_{P} F$ on ${\cal{P}}_{-}$ is identical 
to $F$ on ${\cal{P}}_{+}$. Thus Eq.(\ref{positive}) reads 
\begin{eqnarray}
Z^{-1}  \sum_{\{\vec{S}_{j}\} \in
P}  \Big[ \sum_{\{\vec{S}_{i}\} \in
P_{+}}   \nonumber
\\ F(\{\vec{S}(\vec{x}_{i} \in P_{+})\}, \{\vec{S}(\vec{x}_{j}
\in P)\}) \Big]^{2} \ge 0, 
\end{eqnarray}
which proves the prinicipal Eq.(\ref{basic_ineq}).
Inserting
${\cal{O}} = A - B \langle B R_{p} A \rangle/\langle B R_{p} B \rangle$
in Eq.(\ref{basic_ineq}),
a Schwarz inequality 
\begin{eqnarray}
\langle A R_{P} B \rangle^{2} \le \langle
A R_{P} A \rangle \times \langle B R_{P}B \rangle
\label{schwarz}
\end{eqnarray}
results.
Setting ${\cal{O}} = A - \langle A
\rangle$ in Eq.(\ref{basic_ineq}) (or $B=1$)
\begin{eqnarray}
\langle A \rangle ^{2}
 \le \langle
A R_{P} A \rangle.
\label{Gibbs}
\end{eqnarray} 
We note \cite{oldme} that this general inequality
allows us immediately prove that 
in symmetric (Reflection Positive), opposite
{\it topological charges always
attract} \cite{Opposite}- a similar
yet weaker version
of the relations
derived in \cite{Nussinov-92}.
This provably attractive nature in symmetric systems \cite{oldme}
has ramifications in numerous arenas, including the sign of the Casimir forces \cite{klich}. 
We may repeat the last step
$n$ times in a row 
to get 
\begin{eqnarray}
\langle A \rangle ^{2^{n}} \le  \langle A(R_{P_{1}} A(R_{P_{2}} A(
... R_{P_{n}} A ) ...)) \rangle. 
\label{repeat}
\end{eqnarray}
If we set the size of the system
to be  $N=2^{n}$ then the 
last inequality 
would imply
that 
\begin{eqnarray}
\langle local~ A \rangle \le  \langle Reflected~A \rangle ^{(W/N)}
\label{chessboard}
\end{eqnarray}
where $local~A$ denotes the local 
function $A$ and $Reflected~A$ denotes
the function $A$ generated by consecutive 
reflections of $A$ in different
planes, until it covers the
entire lattice. We have more cautiously
inserted a constant $W$ instead
of one. The functions $A$ that were of 
interest to us were local test functions
that were equal to 
one if a specific local 
configuration was found
and were zero  otherwise.
In this manner 
the probability of finding 
certain necessarily ``bad'' configurations,
lying on the interface between 
the two ground state 
domains, may be bounded.
Here the thermal average 
$\langle A \rangle$ 
is the probability that 
such a ``bad'' configuration would
be found. 
\begin{eqnarray}
\langle Reflected~A \rangle = \frac{Z_{Reflected~A}}{Z}
\label{weight}
\end{eqnarray}
where $Z_{Reflected~A}$ is the sum
of all Boltzmann weights corresponding
the Reflected A configuration which spans the 
entire lattice, and $Z$ is, as usual, 
the sum of all Boltzmann weights- the 
partition function. Eqs. (\ref{chessboard}) and (\ref{weight})
bound the probabilities of certain local configurations.

We will now fortify and extend 
these relations to 
reach the more sophisticated 
inequality (Eq.(\ref{multi})) 
employed in our proof.
Suppose that $A_{1}$ denotes the
observation of a certain local configuration
about point $\vec{x}_{1}$ while $A_{2}$
denotes the test function for another
configuration being seen at $\vec{x}_{2}$.
By the Schwarz inequality (Eq.(\ref{schwarz}))
\begin{eqnarray}
\langle A_{1} A_{2} \rangle ^{2} = \langle A_{1} R_{p}
\overline{A}_{2} 
\rangle ^{2} 
\le \langle A_{1} R_{p} A_{1} \rangle  \times \langle A_{2} R_{p}
A_{2} \rangle,
\label{two}
\end{eqnarray}
where $\overline{A} \equiv R_{p} A$.
We now repeatedly apply Eq.(\ref{Gibbs}) 
(as in Eq.(\ref{repeat}))
to the right hand side of Eq.(\ref{two}) to obtain
\begin{eqnarray}
\langle A_{1} A_{2} \rangle  \le   \langle Reflected~A_{1} \rangle
^{(W/N)} \nonumber
\\
\times 
 \langle Reflected~A_{2} \rangle ^{(W/N)},
\label{TWO}
\end{eqnarray}
which may be followed by an insertion of 
the identity in Eq.(\ref{weight}).
For the correlation function of three local configuration test
functions we may note
that
\begin{eqnarray} 
\langle A_{1} A_{2} A_{3} \rangle ^{3}  = \langle (A_{1} A_{2}) R_{p} 
\overline{A}_{3} \rangle \times \nonumber
\\ 
\langle  (A_{1} A_{3}) R_{p} \overline{A}_{2} \rangle 
\times \langle (A_{2} A_{3}) R_{p} \overline{A}_{1} \rangle,
\end{eqnarray}
i.e. a product of terms of the form $\langle A R_{p} B \rangle$.
We may consequently apply the  
Schwarz inequality, as 
was done in Eq.(\ref{two}),
and then invoke Eq.(\ref{weight})
to get a simple 
extension of Eq.(\ref{TWO}).
Thus, as the reader can see, for the $\ell$-th order correlator 
we have
\begin{eqnarray}
\langle \prod_{i=1}^{\ell} A_{i} \rangle \le  \prod_{i=1}^{\ell}
\Big[ \frac{Z_{Reflected~A_{i}}}{Z} \Big]^{W/N}.
\label{multi}
\end{eqnarray}
Eqs.(\ref{multi}) are a
central result.
With these relations 
(the {\em chessboard estimates}), local 
correlations (appearing on 
the left-hand side)
may be efficiently bounded
in terms of global events
(which form  the 
right-hand side).

\section{Appendix B: Contour Arguments}

In a two dimensional nearest neighbor Ising system 
the probability of having a certain
specific contour of length $\ell$
separating an island of ``-'' ($``N_{G_{1}}''$) 
spins in a sea of ``+'' spins
(or vice versa) is 
bounded by $\exp[ - 2 \beta J \ell]$ 
when the exchange constant units 
are restored. Here we repeat a variant of
the standard Peierls argument for 
the benefit of our readers. 
Let us assume that there are $``+''$ conditions
at infinity. We may evaluate the probability
that the lattice site $\vec{x} = 0$ is occupied 
by a $``-''$ spin. To have a minus
spin at the origin there must be at least one contour
$\Gamma$ encircling the site $0$.

\begin{eqnarray}
Prob(S(\vec{x}=0)= -1) \le \sum_{\Gamma~ \mbox{enc.}~ 0} Prob(\Gamma) 
\label{roman}
\end{eqnarray}

The probability of a given contour $\Gamma$ of
length $\ell$:
\begin{eqnarray} 
Prob(\Gamma)  = \frac{Z_{\Gamma}}{Z} < \frac{Z_{\Gamma}}{Z_{\Gamma~
\mbox{flipped}}}= \exp[- 2 \beta J \ell],
\label{roman1}
\end{eqnarray}
where $Z_{\Gamma~
\mbox{flipped}}$ is the
sum of Boltzmann weights over 
all spin configurations
in which the contour $\Gamma$
is flipped: All of the spins
inside $\Gamma$ are flipped
with all other spins in the system
are untouched. The energy
of each individual
configuration summed upon
in $Z_{\Gamma}$ is
exactly $2 J \ell$
higher than that
configuration
in which 
$\Gamma$ is flipped-
whence the second equality in 
Eq.(\ref{roman1})
follows.  A moment's reflection reveals 
that $Z_{\Gamma~ \mbox{flipped}}$ only contains 
a subset of all the terms which 
are present in $Z$ and consequently 
$Z_{\Gamma~
\mbox{flipped}} < Z$, implying the 
inequality in Eq.(\ref{roman1}).
Eq.(\ref{roman})
implies that
\begin{eqnarray}
Prob(S(\vec{x}=0)= -1) < \sum_{\ell=4,6, \ldots}  n(\ell) 
\exp[- 2 \beta J \ell].
\end{eqnarray}
We now bound the number, $n(\ell)$, 
of contours of specified length $\ell$. 
This number is bounded from above 
by the number of 
unrestricted random walks
of length $\ell$. If given a unique
starting point, the number of
non-backtracking random walks
of length $\ell$ 
on the square lattice is bounded
by $4 \times 3 ^{\ell-1}$.
The number of possible starting points 
for a random walk enclosing the 
origin ($\vec{x}=0$) is strongly bounded
by the area enclosed by the
contour.
For a given contour length $\ell$ this
maximal area is
$(\ell/4)^{2}$, 
corresponding to a square 
of side $(\ell/4)$. Bounding 
this maximal area by $(3 \ell^{2}/4)$
provides a very generous 
over estimate.

Fusing all of the bounds 
together:
\begin{eqnarray}
Prob(S(\vec{x}=0)=-1) =  \langle \frac{N_{G_{1}}}{N}  
\rangle < \nonumber
\\  
\sum_{\ell=4,6,...} \ell^{2}
3^{\ell} \exp[-2 \beta J \ell].
\label{standard}
\end{eqnarray} 
The last sum can be made 
smaller than 1/2
(indicating symmetry 
breaking) for sufficiently 
large $\beta$ (low 
temperatures). The proofs presented 
for the XY models 
(subsection(\ref{RP}))
rest on the same logic
with the twist that
the contour inversion 
probability bounds  (Eq.(\ref{roman1}))
are replaced 
by the Reflection Positivity 
inequality of Eq.(\ref{multi}).  
In \cite{clock}, similar bounds were employed to prove
a low temperature orders;  exact duality arguments
were then invoked to establish the existence 
of an upper Kosterlitz-Thouless type transition 
temperature for $Z_{k>4}$ models. As noted earlier,
in the $k \to \infty$ limit, $Z_{k} \to O(2)$ which cannot be broken
in two-dimensions and only the upper (Kosterlitz-Thouless) transition temperature
remains.


\begin{thebibliography}{10}
\bibitem{continuum}

In a continuous quantum
system, the amplitude of
the wavefunction
$|\psi(\vec{x})|$ vanishes
at all singular points (e.g. vortices)
to effectively excise
these points so that a
wave function 
may be globally defined.
For the most part, we will consider 
an effective underlying 
lattice on which
the modulus of the 
wave function
is constant;
the plaquette 
``vortices'' 
fall in between 
the lattice sites.
There are no true 
singularities
for such lattice
systems.
\bibitem{degs_of_freedom}
Note that there are $2N$ ~ link gauges $\{A_{ij}\}$s
(in a system with periodic boundary conditions).
As noted earlier, in a system with open boundary
conditions there are $(2L^{2}+2L)$ bond angles 
$\{\phi_{ij}\}$. Subtracting from
this quantity the total number of 
constraints (of net circulation
about each plaquette), $N=L^{2}$,
we are left with $[(L+1)^2-1]$
degrees of freedom.
The latter quantity can be interpreted as
the total number of spins minus one 
global rotation. The spectrum  (Eq.(\ref{definition})) 
is a continuous function of these.
\bibitem{modular}
For a system with periodic boundary
conditions, there
are $N=L^{2}$ circulation constraints
of the form in Eq.(\ref{f})
and $2L^{2}$ gauge 
variables $\{A_{ij}\}$.
This leaves $N$ 
independent gauge variables
 $\{A_{ij}\}$. 
Now imagine fixing 
$N$ arbitrary $\{A_{ij}\}$
such that no three
variables are common
to the same plaquette.
After such a gauge fixing
procedure, the remaining $N$ 
dependent gauge link
variables are fixed by the 
circulation sums of Eq.(\ref{f}).
Now consider shifting the flux
threading each plaquette 
$\Phi \to \Phi + n \Phi_{0}$,
or, equivalently, the frustration 
$f \to f + 2 \pi n$,
with an integer $n$.
This amounts, in a given gauge, 
to changing the $N$ dependent
gauge link variables by an integer multiple of 
$2 \pi$. By the periodicity 
of the cosine interaction,
the spectrum will 
remain trivially invariant.
\bibitem{screening}
The following is standard.
The kinetic energy
\begin{eqnarray}
E_{kin} = \frac{1}{2} \int \rho(\vec{x})v^{2}(\vec{x}) d^{2}x  \nonumber
\\ = \frac{1}{8 \pi} \int \frac{mc^{2}}{4 \pi n e^{2}}(\nabla \times H)^{2} d^{2}x,
\end{eqnarray}
where we invoked Ampere's
law $\nabla \times H = \frac{4 \pi}{c} \vec{j}$.
The energy (kinetic + field) of a vortex line of length
$L$ 
\begin{eqnarray}
E = \frac{L}{8 \pi} \int d^{2}x~[H^{2}(\vec{x})+ \lambda_{L}^{2}(\nabla \times H)^{2}]
\label{tot_energy}
\end{eqnarray}
where 
$\lambda_{L}^{2}=mc^{2}/(4 \pi n e^{2})$ marks
the ratio between the kinetic energy
and the magnetic energy. The resulting variational
(London) Eq. for the field configuration reads
\begin{eqnarray}
\vec{H} + \lambda_{L}^{2} \nabla \times (\nabla \times \vec{H}) = 0 
\end{eqnarray}
A vortex may be described by a 
source term,
\begin{eqnarray}
\vec{H} + \lambda_{L}^{2} \nabla \times \nabla \times H = q_{0} \hat{z} \delta^{2}(\vec{x})
\end{eqnarray}
with $q_{0}$ the flux.
In cylindrical coordinates
\begin{eqnarray}
H_{z}-\frac{\lambda_{L}^{2}}{r} \frac{d}{dr}(r \frac{dH_{z}}{dr}) = q_{0} \delta^2(\vec{x}).
\end{eqnarray}
The solution with the required singular behavior
near the origin is
\begin{eqnarray}
H_{z} = \frac{q_{0}}{2 \pi \lambda_{L}^{2}}K_{0}(\frac{r}{\lambda_{L}}),
\end{eqnarray} 
with $K_{0}$ a Bessel function.
If there are two vortices then on the right hand side
of the London equation we should insert 
two source terms: $\hat{z} [q_{1} \delta^{2}(\vec{x}-\vec{x}_{1})+ q_{2} \delta^{2}(\vec{x}-\vec{x}_{2}]$. The corresponding solution is 
a simple super-position,
\begin{eqnarray}
H_{z} = \frac{1}{2 \pi \lambda_{L}^{2}}[q_{1} K_{0}(\frac{|\vec{x}-\vec{x}_{1}|}{\lambda_{L}})+ q_{2} K_{0}(\frac{|\vec{x}-\vec{x}_{2}|}{\lambda_{L}})].
\end{eqnarray}
Inserting this in Eq.(\ref{tot_energy}), 
the interaction energy is seen to be
\begin{eqnarray}
\frac{E^{12}}{L} = 
 2 (\frac{q_{1}q_{2}}{(4 \pi \lambda_{L})^{2}})K_{0}(\frac{|\vec{x}_{1}-\vec{x}_{2}|}{\lambda_{L}}).
\end{eqnarray} 
Asymptotically,
\begin{eqnarray}
K_{0}(x) \sim \ln (\frac{1}{x})~~~~ (x \ll 1) \nonumber
\\ K_{0}(x) \sim \sqrt{\frac{\pi}{2x}} \exp[-x] ~~~ (x \gg 1).
\label{asym}
\end{eqnarray}
$K_{0}(x)$ is monotonically decreasing with
$x$ leading, at all distances, to an attractive/repulsive forces 
between opposite/identical poles.
The top (small $x$) form of Eq.(\ref{asym}) 
is the one encountered in this work. 
We identify the length scale appearing
in the argument of the logarithm of our 
continuum limit treatment with the 
screening length $\lambda_{L}$- tending 
to the system size in 
the absence of screening.
\bibitem{Laugh}
We also note that at a complete
extreme- if the requirement electronic 
quantum mechanical wavefunction have a fixed
amplitude $|\psi(\vec{x})|=1$
 were relaxed then instead of a phase only 
model we would arrive, in the continuum limit
(replacing the cosine by a covariant gradient $D_{\mu} =
\partial_{\mu} - i (e^{*}/c)A_{\mu}$),
at the Quantum Hall problem.

The Laughlin like wavefunction that we found in Eq.(\ref{Laugh}) 
is to be expected. $\psi(z)$ was a function of a single 
spinless electronic coordinate the in the background of 
topological charges incurred
by the other (formally ``frozen'') 
electrons. (This is, of course,
merely formal and not a valid physical 
procedure for the system of 
identical particles.) The charge $q$
plays the role of the inverse filling
fraction $\nu^{-1}$. If the 
amplitude of the wavefunction
is allowed to vary (or, equivalently,
if the phase $\phi_{ij}$ is allowed
to become complex subject only
a spherical (large $n$) constraint
$\langle |\phi_{ij}|^{2} \rangle =1$)
then the problem becomes Gaussian.
Alexander \cite{Alexander}
has shown that in this approximation 
the system is described by the 
Harper Eq. \cite{Hofstadter}, \cite{Harper}.
\bibitem{Non-Abelian}

As mentioned earlier, one way of 
describing a uniformly frustrated spin model 
is in terms of parallel transport 
on a sphere. Consider a three-spin 
$\vec{S}(\vec{x})$ with the 
\begin{eqnarray}
H = -\sum_{\langle \vec{x},\vec{y} \rangle} \vec{S}(\vec{x}) \circ \vec{S}(\vec{y})
\end{eqnarray}
where the $\circ$ operation denotes the scalar
product between a spin $\vec{S}(\vec{x})$ rolled on a sphere
of radius $R = \theta^{-1} \sim f^{-1/2}$ to
a neighboring site $\vec{y}$ and the spin $\vec{S}(\vec{y})$
situated at $\vec{y}$. 
If the frustration is uniform 
then, in momentum space, a viable representation is
\begin{eqnarray}
H = \frac{1}{2N} \sum_{\vec{k}} S_{i}(\vec{k}) 
v_{ij}(\vec{k}) S_{j}(-\vec{k})
\end{eqnarray}
where 
\begin{displaymath}
\mathbf{v_{ij}(\vec{k})} = -  \left (\begin{array}{ccc}
 A  &  0 & -D \\
0 &  B & -E \\\
D &  E & C
\end{array} \right)
\end{displaymath}    
with
\begin{eqnarray}
A = \cos \theta \cos k_{1} + \cos k_{2}, ~ B = \cos k_{1} + \cos k_{2} \cos \theta \nonumber
\\ C = \cos \theta \sum_{\l=1}^{2} \cos k_{l},  \nonumber 
\\ D = i \sin \theta \sin k_{1}, ~ E = i \sin \theta \sin k_{2}. 
\end{eqnarray}
Locally, within the tangential spin coordinate system of
the sphere, the spin has two components.
\bibitem{pv}
For infinitesimal $\Delta E$, 
a simple calculation reveals that
the fraction of 
unfrustrated ($f=0$) XY states having energies between
the ground state energy $E_{ground}$ 
and $E_{ground} + \Delta E$
relative to the net ``phase space''
volume $(2 \pi)^{N}$
is trivially
$fraction~ \approx (\frac{\Delta E}{2 \pi J})^{N/2} 
\frac{1}{\Gamma(N/2+1)}$, 
when dimensions are fully restored.
\bibitem{lat}  Note that if this plaquette was immersed in a lattice in 
which all other spins $i$ apart from
those belonging to our plaquette point 
in the same direction ($\theta_{i} = const$)
then an antivortex of strength $m= -1$ 
would appear immediately adjacent
to it. Vortex neutrality 
is strictly enforced in any system
with periodic boundary conditions
(of which the condition $\theta_{i} = const$  
is one extreme example allowing 
a $3 \times 3$ plaquette system to be 
wrapped around a
two-torus).
\bibitem{sophisticated} Though not imperative here,
more sophisticated bounds may be enacted via 
spin-wave calculations. See, e.g., \cite{BCN},  Lemma 6.8 in particular.
\bibitem{II}

Further, as the magnetic
flux $f$ piercing a type II 
superconductor increases,
the superconducting $T_{c}$ 
(associated with phase 
ordering) monotonically 
decreases. By a naive analogy one 
might expect the phase 
``ordering'' temperature 
$T_{KT}(f)$ to drop monotonically 
for small $f$.
\bibitem{construction}
If any Hamiltonian is initially defined only on a finite 
size fragment of the lattice (where it is $B$) then by applying 
Eq.(\ref{rp-symm}) repeatedly 
with different bounding planes $P_{\alpha}$, we may generate 
a reflection positive Hamiltonian 
(with respect to the planes $\{P_{\alpha}\}$) 
which spans the entire 
lattice.
\bibitem{Opposite} 
The probability of having a vortex-antivortex 
pair is greater than or equal to 
having vortices separated by an infinite 
distance. The proof is trivial:
Let $A$ denotes the event of 
finding a vortex centered about 
point $\vec{x}$: $A=1$ if a vortex is
found at $\vec{x}$ and is zero otherwise.
In such a case $\langle A R_{p} A \rangle$ denotes the
probability of finding one vortex
at $\vec{x}$ and one antivortex at
its mirror image $R_{p} \vec{x}$. 

The inequality derived in Appendix A (Eq.(\ref{Gibbs})) 

\begin{eqnarray}
\langle A R_{p} A \rangle \ge \langle A \rangle^{2}
\end{eqnarray}
reads, upon letting $A$ be the above function,
\begin{eqnarray}
Prob(vortex~at~\vec{x}~ and ~antivortex~ at ~R_{p} \vec{x}) 
\nonumber \\ \ge
[Prob(vortex~ at ~\vec{x})]^{2}.
\end{eqnarray}

$[- \ln \langle A R_{p} A \rangle]$ is, diagrammatically,
the free energy of the system with two additional 
external legs at $\vec{x}$ and $R_{p} \vec{x}$ minus
the free energy of the closed system 
without any external topological 
charges. The last set of inequalities implies that
the free energy of a finite separation 
vortex-antivortex pair is lower than 
that at infinite distance. This also applies to situations wherein
uniform boundary effects are considered- these
may be reflected relative to a plane
$P$ orthogonal to the boundary. This applies
to all topological
charges in
Reflection Positive
theories.
\end{thebibliography}
\end{document}